\documentclass[aps,prd,nofootinbib,twocolumn,superscriptaddress,preprintnumbers,balancelastpage,longbibliography]{revtex4-1}

\usepackage{appendix}
\usepackage[utf8]{inputenc}
\usepackage{amsmath,amssymb,mathtools,bm}
\usepackage{graphicx, color, hepunits}
\usepackage[dvipsnames]{xcolor}
\usepackage{float}
\usepackage{fontawesome5}
\usepackage{multirow}
\usepackage{tikz-feynman} 
 \usepackage{hyperref}
\hypersetup{
    colorlinks=true,       
    linkcolor=blue,        
    citecolor=blue,        
    filecolor=magenta,     
    urlcolor=blue          
}
\usepackage[utf8]{inputenc}
\usepackage[english]{babel}
\let\vec\mathbf
\usepackage{tensor}
\usepackage{xspace} 
\usepackage{physics} 
\usepackage{soul} 

\usepackage{pifont}

\usepackage{url}
\usepackage{appendix}
\usepackage{amsmath,amssymb,mathtools,bm}
\usepackage{graphicx, color, hepunits}
\usepackage{float}
\usepackage{multirow}
\usepackage{filecontents}

\usepackage{tabularx}
\usepackage{array}
\newcolumntype{P}[1]{>{\centering\arraybackslash}p{#1}}
\newcommand{\ra}[1]{\renewcommand{\arraystretch}{#1}}

\makeatletter
\def\hlinewd#1{%
\noalign{\ifnum0=`}\fi\hrule \@height #1 \futurelet
\reserved@a\@xhline}
\makeatother
\newcolumntype{L}[1]{>{\raggedright\let\newline\\\arraybackslash\hspace{0pt}}m{#1}}
\newcolumntype{C}[1]{>{\centering\let\newline\\\arraybackslash\hspace{0pt}}m{#1}}
\newcolumntype{R}[1]{>{\raggedleft\let\newline\\\arraybackslash\hspace{0pt}}m{#1}}
\renewcommand{\arraystretch}{1.2}

\definecolor{overleaf}{rgb}{0.0, 0.7, 0.0}

\renewcommand{\eqref}[1]{Eq.~(\ref{#1})}
\newcommand{\bbm}{\begin{bmatrix}}
\newcommand{\ebm}{\end{bmatrix}}

\newcommand{\expn}[1]{\left\langle#1\right\rangle}
\newcommand{\vx}{\vec{x}}

\newcommand{\xray}{\textit{X}-ray\xspace}

\newcommand{\cmfast}{\texttt{21cmFAST}\xspace}
\newcommand{\dmcm}{\texttt{DM21cm}\xspace}
\newcommand{\dhis}{\texttt{DarkHistory}\xspace}

\begin{document}

\title{Inhomogeneous Energy Injection in the 21-cm Power Spectrum: \\ Sensitivity to Dark Matter Decay}

\author{Yitian Sun}
\email{yitians@mit.edu}
\affiliation{Center for Theoretical Physics, Massachusetts Institute of Technology, Cambridge, Massachusetts 02139, U.S.A}

\author{Joshua W. Foster}
\email{jwfoster@mit.edu}
\affiliation{Center for Theoretical Physics, Massachusetts Institute of Technology, Cambridge, Massachusetts 02139, U.S.A}

\author{Hongwan Liu}
\affiliation{Kavli Institute for Cosmological Physics, University of Chicago, Chicago, IL 60637}
\affiliation{Theoretical Physics Department, Fermi National Accelerator Laboratory, Batavia, IL 60510}

\author{Julian B.~Mu\~{n}oz}
\affiliation{Department of Astronomy, The University of Texas at Austin, 2515 Speedway, Stop C1400, Austin, TX 78712, USA}

\author{Tracy R.~Slatyer}
\affiliation{Center for Theoretical Physics, Massachusetts Institute of Technology, Cambridge, Massachusetts 02139, U.S.A}

\date{\today}
\preprint{MIT-CTP/5657}
\preprint{FERMILAB-PUB-23-0816-T-V}

\begin{abstract}
The 21-cm signal provides a novel avenue to measure the thermal state of the universe during cosmic dawn and reionization (redshifts $z\sim 5-30$), and thus to probe energy injection from decaying or annihilating dark matter (DM). 
These DM processes are inherently inhomogeneous: both decay and annihilation are density dependent, and furthermore the fraction of injected energy that is deposited at each point depends on the gas ionization and density, leading to further anisotropies in absorption and propagation. In this work, we develop a new framework for modeling the impact of spatially inhomogeneous energy injection and deposition during cosmic dawn, accounting for ionization and baryon density dependence, as well as the attenuation of propagating photons. We showcase how this first complete inhomogeneous treatment affects the predicted 21-cm power spectrum in the presence of exotic sources of energy injection, and forecast the constraints that upcoming HERA measurements of the 21-cm power spectrum will set on DM decays to photons and to electron/positron pairs. These projected constraints considerably surpass those derived from CMB and Lyman-$\alpha$ measurements, and for decays to electron/positron pairs they exceed all existing constraints in the sub-GeV mass range, reaching lifetimes of $\sim 10^{28}\text{ s}$. Our analysis demonstrates the unprecedented sensitivity of 21-cm cosmology to exotic sources of energy injection during the cosmic dark ages. Our code, \texttt{DM21cm}, includes all these effects and is publicly available in an accompanying release \href{https://github.com/yitiansun/DM21cm}{\faGithub}.
\end{abstract}
\maketitle

\section{Introduction}

The redshifted 21-cm signal produced by the hyperfine transition of neutral hydrogen represents the leading prospect for studying cosmology at intermediate redshifts between the CMB formation at early times and late-time large-scale structure surveys. Measurements of the 21-cm signal are expected to provide a window into the thermal and ionization history of our universe between the end of the cosmic dark ages and reionization. Experiments such as EDGES~\cite{Monsalve:2016xbk}, LEDA~\cite{Spinelli:2022xra}, PRI$^\mathcal{Z}$M~\cite{prizm}, and SARAS~\cite{Singh:2017syr} are already in the process of measuring the global (monopole) 21-cm signal as a function of frequency.
Additionally, radio interferometers like PAPER~\cite{Pober:2013ig}, the MWA~\cite{Tingay:2012qe}, LOFAR~\cite{Rottgering:2003jh}, HERA~\cite{DeBoer:2016tnn}, and the upcoming Square Kilometre Array (SKA)~\cite{Dewdney:2009tmd}, will measure the frequency-dependent spatial variations of the 21-cm signal.
Current power-spectrum limits are cutting into physically motivated parameter space~\cite{gharalofar, Trott:2020szf, HERA:2021noe, HERA:2022wmy}, and upcoming observations are expected to reach deep enough to detect the reionization signal.

In $\Lambda$CDM cosmology, reionization is expected to be driven by the star formation within the first galaxies. The first stars emit ionizing radiation, creating patches of fully ionized hydrogen which grow to fill the universe, leading to the present-day fully ionized intergalactic medium (IGM). However, more exotic sources of energy injection such as the annihilation or decay of massive dark matter (DM) particles to energetic,  electromagnetically interacting particles could play an important and even detectable role in the process of reionization (including changes to the ionization and thermal history well before the universe fully reionizes). In the DM scenario, the reionization process may depart significantly from the standard astrophysical scenario; the emission of radiation that ionizes and heats the universe will now partly track the spatial distribution of DM rather than the stellar distribution, and will occur on a timescale set by the DM depletion mechanism rather than the star formation rate (SFR). The deviations are expected to be imprinted in the 21-cm signal much as they might be in the CMB, a scenario which has been well studied in, \textit{e.g.}, Refs.~\cite{Adams:1998nr, Chen:2003gz, Padmanabhan:2005es, Slatyer:2009yq, Chluba:2011hw, Weniger:2013hja, Galli:2013dna, Slatyer:2016qyl}. The CMB and 21-cm signal measurements are expected to be highly complementary, with the CMB providing good sensitivity to DM energy injection at early times, such as through $s$-wave annihilation, while the 21-cm signal should produce improved constraints on scenarios where the energy injection is weighted more toward later times,  such as through decay or $p$-wave annihilation.

The most sensitive possible 21-cm probes of annihilating or decaying DM will make use of both high precision monopole measurements and measurements of the power spectrum in joint analyses making use of data collected across a range of observatories and facilities. While the effect of dark matter on the global monopole signal has been previously studied in several contexts (see Refs.~\cite{2013MNRAS.429.1705V,Evoli_2014, Clark:2018ghm, Mitridate:2018iag, DAmico:2018sxd, Hektor:2018qqw}), the power spectrum measurement is particularly compelling as the combination of spatial and temporal information can be utilized to break degeneracies with uncertainties in the standard astrophysics. However, at redshifts relevant for the production of measurable 21-cm radiation, the linear and then nonlinear growth of structure has produced a universe which is inhomogeneous at the $\mathcal{O}(1)$ level. Understanding the impact of these inhomogeneities on both the 21-cm global signal and power spectrum is critical for making accurate predictions in the DM paradigm.

\begin{figure}[!t]
    \begin{center}
    \includegraphics[width=0.49\textwidth]{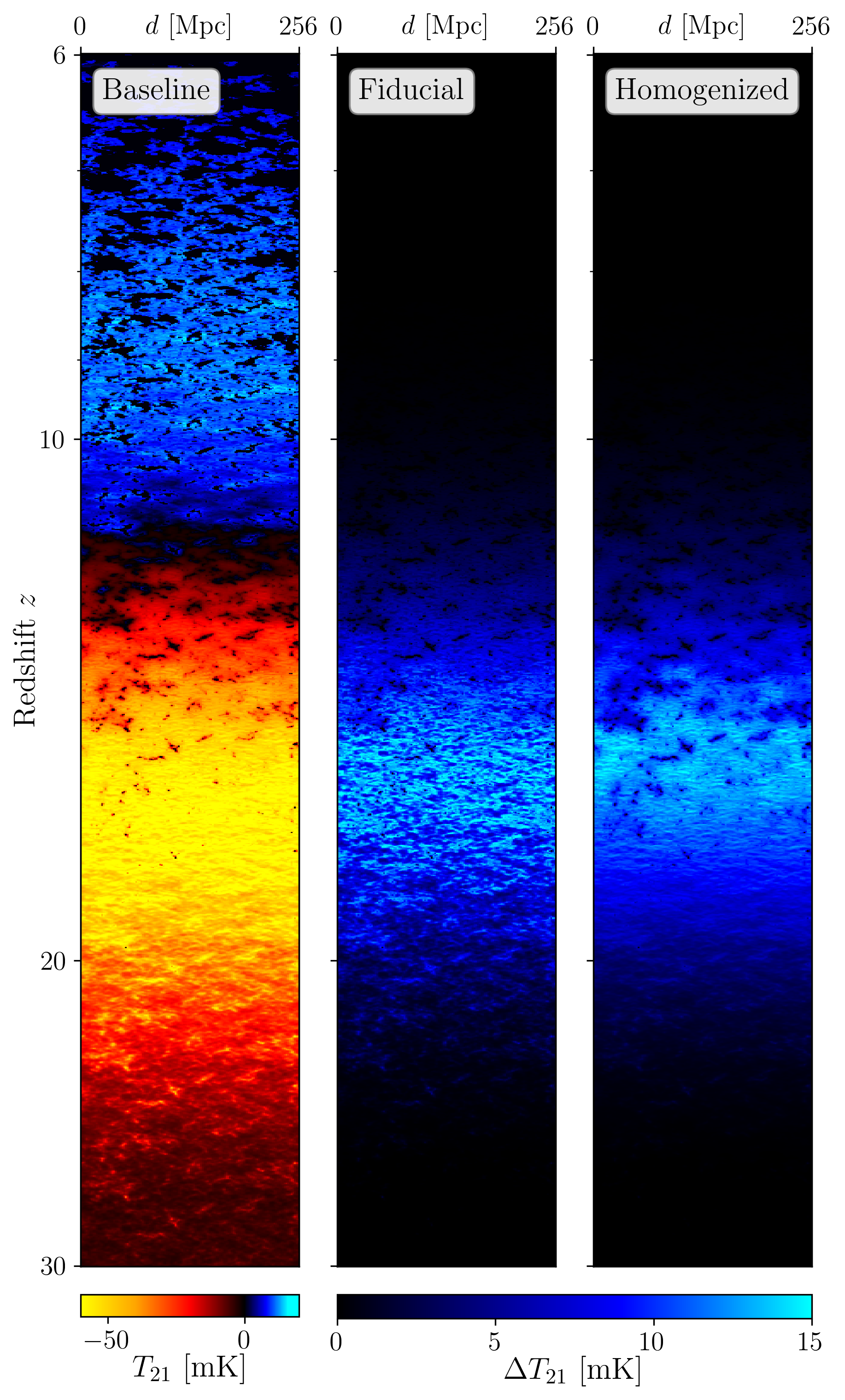}
    \vspace{-0.6cm}
    \caption{
    \textbf{Effects of Inhomogeneous and Homogeneous DM Energy Injection on 21-cm Brightness Temperature.} The left-most lightcone shows the baseline $T_{21}$ in which reionization is driven by standard astrophysical processes with no exotic energy injection from DM. The center and the right lightcones show the change in $T_{21}$ from the baseline scenario in the presence of  $1\,\mathrm{keV}$ mass DM decaying monochromatically to two photons with a lifetime of  $10^{28.8}$~s, which is currently allowed but we will show can be tested by upcoming 21-cm data. The difference in the middle lightcone is obtained by treating the exotic energy injection of DM with our new spatially inhomogeneous procedure, while the right lightcone is obtained from a simpler homogenized procedure that has been considered in previous literature, \textit{e.g.}, Ref.~\cite{Facchinetti:2023slb}. While the sensitivity of HERA under the simplified homogenized procedure is ultimately similar to the full calculation, they produce very distinct shifts from the baseline $T_{21}$ evolution.}
    \label{fig:lightconediff}
    \end{center}
    \vspace{-0.3cm}
\end{figure}

A similar challenge exists even in the standard stellar reionization scenario due to the considerable spatial inhomogeneities in the star formation rate density (SFRD), which has been studied in large-scale radiation-magnetohydrodynamic simulations, such as~\cite{Kannan:2021xoz,coda, sphinx}. However, these simulations are costly and often cannot resolve the smallest star-forming galaxies along with the long mean-free paths of \xray radiation. This limits their use in analyses that seek to understand the impact of astrophysical parameters and modeling on the 21-cm signal. As such, semi-numerical simulation frameworks such as \cmfast~\cite{2011MNRAS.411..955M} and alternatives~\cite{Santos:2009zk,Visbal:2012aw,Mirocha:2014faa,Munoz:2023kkg} have become a standard tool for 21-cm analyses, offering a more computationally efficient approach. By default, \cmfast does not include the effects of DM beyond its role in the growth of structure, though some recent efforts have been made to include the effects of dark matter elastic scattering~\cite{Flitter:2023mjj} or the role of exotic energy injections under an approximation of homogeneous energy deposition~\cite{Facchinetti:2023slb}.

In this work, we develop a new simulation framework, which we call \dmcm, that joins the simulation procedure implemented in \cmfast with the ionization and thermal history modeling of the public \dhis code package,\footnote{Throughout this work we employ version 1.1 of \dhis \cite{Liu:2019bbm, Sun:2022djj}. Recent improvements to the code \cite{Liu:2023fgu} have been shown to have only small effects on the cosmic thermal and ionization histories \cite{Liu:2023nct}, although the impact on Lyman-$\alpha$ photons may be more substantial and may merit further study.} in order to perform first-of-their-kind simulations of ionization and heating in the presence of spatially inhomogeneous exotic energy injection and deposition. Though designed with energy injection due to DM in mind, \dmcm is a flexible tool capable of accommodating arbitrary spatial and temporal dependence for the energy injection process. As an illustrative example of the power of our simulation framework, in Fig.~\ref{fig:lightconediff}, we show a predicted change in the 21-cm brightness temperature lightcone including $1\,\mathrm{keV}$ DM decaying to photons with a lifetime of $6\times 10^{28}\,\mathrm{s}$ computed by \dmcm. We also compare this to the prediction made with the simplifying assumption of homogeneous exotic energy injection and deposition, such as was made in Ref.~\cite{Facchinetti:2023slb}. Manifest differences between the predicted spatial morphology of the brightness temperature at redshifts between 10 and 20, corresponding to radio frequencies probed by existing and upcoming observatories, make clear the importance of a modeling procedure that treats inhomogeneities accurately.

This paper is organized as follows. In Sec.~\ref{sec:Basics}, we review the basic aspects of 21-cm cosmology, with a particular focus on the calculations performed by \cmfast that we will extend. In Sec.~\ref{sec:Model}, we detail the modeling prescription and implementation of the \dmcm framework and study monopole and power spectrum signals generated for extreme but instructive decaying DM scenarios. In Sec.~\ref{sec:Proj}, we use our \dmcm framework to perform a Fisher forecast, projecting sensitivities and limits on DM decay across many decades of DM mass for two benchmark scenarios: monochromatic decays to photons and to electron/positron pairs. These projected constraints surpass existing ones from the CMB and Ly$\alpha$ by several orders of magnitude and, when realized, will provide leading sensitivity to sub-keV DM decay to photons and sub-GeV DM decay to electron/positron pairs. Finally, we offer some concluding remarks in Sec.~\ref{sec:Conclusion}, with some numerical tests and systematic variations presented in the Appendices.

\section{Review of 21-cm Cosmology and \cmfast}
\label{sec:Basics}

We begin with a brief review of the most relevant aspects of 21-cm cosmology; for a detailed review of this field we refer the reader to e.g.\ Ref.~\cite{Pritchard_2012}. The fundamental observable associated with the 21-cm signal is the brightness temperature of the redshifted 21-cm line relative to the CMB blackbody temperature. This is a frequency-dependent line-of-sight quantity which is given by 
\begin{equation}
\begin{split}
    T_{21}(\nu) &\approx 27 x_\mathrm{HI}(1+\delta)\left( \frac{H}{dv_r/dr + H}\right)\left( 1-\frac{T_\gamma}{T_S}\right) \\
    &\times \left(\frac{1+z}{10} \frac{0.15}{\Omega_M h^2} \right)^{1/2} \left( \frac{\Omega_b h^2}{0.023} \right)\,\mathrm{mK},
\end{split}
\end{equation}
where $\delta$ is the Eulerian density contrast, $H$ is the Hubble parameter, and $dv_r/dr$ is the comoving gradient of the comoving velocity projected along the line of sight. $T_\gamma$ is the CMB temperature, $\Omega_M$ and $\Omega_b$ the respective present-day matter and baryon abundances relative to the critical density, and $h$ the present-day Hubble parameter in units of $100\,\mathrm{km/s/Mpc}$ \cite{Furlanetto:2006tf}. All of these quantities
are independent of any energy injection, which enters the brightness temperature through its effects on the neutral fraction of hydrogen $x_\mathrm{HI}$ and the gas spin temperature $T_S$ \cite{Furlanetto:2006jb}. These quantities are defined in further detail below.

In this section, we review at a general level how these quantities are calculated in \cmfast as well as how they can be modified to account for exotic energy injection, building on the previous treatment described in e.g.~Ref~\cite{Lopez-Honorez:2016sur}.

\subsection{Spin Temperature Evolution}
We begin with a review of the evolution of the spin temperature $T_S$, which defines the occupation level of the triplet excited state with respect to the singlet ground state in the hyperfine two-level system. $T_S$ is jointly determined by hyperfine transitions due to \textit{i)} the absorption and emission of CMB photons, \textit{ii)} collisions between hydrogen atoms and other hydrogen atoms, free electrons, and free protons, and \textit{iii)} the absorption and emission of Lyman-$\alpha$ (Ly$\alpha$) photons, also known as the Wouthuysen-Field effect \cite{1952AJ.....57R..31W, 1959ApJ...129..536F}. The spin temperature can be written as 
\begin{equation}
    T_S^{-1} = \frac{T_\gamma^{-1} + x_c T_k^{-1} + x_\alpha T_\alpha^{-1}}{1 + x_c + x_\alpha},
\end{equation}
where $T_k$ is the gas kinetic temperature and $T_\alpha$ is the effective color temperature of the Ly$\alpha$ radiation field; $x_c$ and $x_\alpha$ are coupling coefficients for the collision and Ly$\alpha$ scatterings~\cite{Furlanetto:2006jb}.
To a good approximation, $T_k \approx T_\alpha$~\cite{1959ApJ...129..536F}, though there exist more precise treatments~\cite{2006MNRAS.367..259H}, and $x_c$ has been calculated in detail by \cite{2005ApJ...622.1356Z, Furlanetto:2006su}. 
It is clear, then, that the thermal state of the IGM will become imprinted onto the spin temperature, and DM energy injection will alter the kinetic temperature $T_k$ and the Wouthuysen-Field coupling coefficient $x_\alpha$.
Let us describe how we compute each of these terms.

\subsubsection{Kinetic Temperature Evolution}

Following the \cmfast treatment detailed in Ref.~\cite{2011MNRAS.411..955M}, the dynamics of the spin temperature evolution are governed by the system (in the absence of any exotic sources of energy injection)
\begin{equation} \label{eq:dxedzp-dTdzp}
\begin{split}
\frac{d x_e(z, \mathbf{x})}{dz} &= \frac{dt}{dz} \left[\Lambda_\mathrm{ion} - \alpha_A C x^2_e n_A f_\mathrm{H} \right] \\
\frac{dT_k(z, \mathbf{x})}{dz} &= \frac{2}{3k_B(1+x_e)} \frac{dt}{dz} \sum_p \epsilon_p\\
&+ \frac{2 T_k}{3 n_A}\frac{dn_A}{dz} - \frac{T_k}{1+x_e} \frac{dx_e}{dz},
\end{split}
\end{equation}
where $x_e$ is the local ionized fraction in the ``mostly neutral" IGM as produced by photoionization by \textit{X}-rays, $n_A$ the local physical nuclear density,\footnote{In \cmfast, $n_A$ is referred to as the `baryon number density' when in fact it is the total number density of hydrogen and helium nuclei.}
$\epsilon_p$ the heating/cooling rate per nucleus,
$\Lambda_\mathrm{ion}$ the ionization rate per baryon, $\alpha_A$ the case-A recombination coefficient, $C\equiv\expn{n_e^2}/\expn{n_e}^2$
the free-electron clumping factor, $k_B$ the Boltzmann constant, and $f_\text{H}\equiv n_\text{H}/(n_\text{H}+n_\text{He})$ the hydrogen nucleus number fraction. In this work, we also follow \cmfast in assuming that the abundance of doubly ionized helium is negligible, and that the singly ionized helium fraction $n_\text{HeII}/n_\text{He}$ is always equal to the hydrogen ionized fraction $n_\text{HII}/n_\text{H}$. In the mostly neutral IGM, the ionized fraction $x_e$ is therefore defined as $x_e=n_\text{HI}/n_\text{H}=n_\text{HeII}/n_\text{He}=n_e/n_A$, where $n_e$ is the number density of free electrons.\footnote{Note that this definition is different from the one adopted in \dhis, where $x_e$ is defined as the ratio of the free electron number density and the number density of all hydrogen nuclei, $x_e=n_e/n_\text{H}$. This is also not equal to $n_e/n_\text{A}$, the definition used in \dmcm and \cmfast.} We additionally comment on the two-phase ionization model in Sec.~\ref{sec:NeutralHydrogenEvolution}. 
In \cmfast, all terms $n_A$, $T_k$, $x_e$, $\Lambda_\mathrm{ion}$ and $\epsilon_p$ are treated as spatially dependent to some limited degree. Though there are no explicit diffusion terms, the ionization and heating rates $\Lambda_\mathrm{ion}$ and $\epsilon_p$ are calculated accounting for radiation which is emitted at time $z'$ from location  $\mathbf{x}'$ and propagates to  $\mathbf{x}$ by $z$. We will discuss this procedure at greater length in Sec.~\ref{sec:Implementation}. Otherwise, the ionization fraction $x_e$ and the kinetic temperature $T_k$ evolve independently at each $\mathbf{x}$. 

In the presence of exotic energy injection such as from DM, consistent with the modeling of \cmfast, we may add two additional terms to these expressions (marked in red)
\begin{equation}
\begin{split}
\label{Eq:ODE}
\frac{d x_e(z, \mathbf{x})}{dz} &= \textcolor{red}{\frac{dx_e^\mathrm{DM}}{dz}}+\frac{dt}{dz} \left[\Lambda_\mathrm{ion} - \alpha_A C x^2_e n_A f_\mathrm{H} \right] \\
\frac{dT_k(z, \mathbf{x})}{dz} &= \textcolor{red}{\frac{dT_k^\mathrm{DM}}{dz}} + \frac{2}{3k_B(1+x_e)} \frac{dt}{dz} \sum \epsilon_p\\
&+ \frac{2 T_k}{3 n_A}\frac{dn_A}{dz} - \frac{T_k}{1+x_e} \frac{dx_e}{dz}.
\end{split}
\end{equation}
Like their standard astrophysics analogs in $\Lambda_\mathrm{ion}$ and $\epsilon_p$, these new terms $\textcolor{red}{dx^\mathrm{DM}_e/dz}$ and \textcolor{red}{$dT^\mathrm{DM}_k/dz$}, will be calculated in a manner that accounts for the spatial dependency of emission and absorption.

\subsubsection{Wouthuysen–Field Coupling}
Like the kinetic temperature, the Wouthuysen–Field coupling $x_\alpha$ also receives a contribution for any exotic energy injection involved in the cosmology. Specifically, this dimensionless parameter is given by
\begin{equation}
    \label{Eq:x_alpha}
    x_\alpha =  1.7 \times 10^{11} \frac{S_\alpha}{1+z} \left(\frac{J_\alpha}{\mathrm{cm}^{-2} \, \mathrm{s}^{-1}\, \mathrm{Hz}^{-1}\mathrm{sr}^{-1} } \right),
\end{equation}
where $S_\alpha$ is an atomic physics correction factor \cite{2006MNRAS.367..259H} calculated in a spatially-dependent way and $J_\alpha$ is the Ly$\alpha$ background intensity. Maintaining the treatment of \cmfast, the calculation of $S_\alpha$ and $T_\alpha$ from atomic physics modeling needs no modification in the presence of exotic energy injection. Rather, the effect of exotic energy injection is accommodated by the minimal modification 
\begin{equation*}
J_\alpha \to J_\alpha + \textcolor{red}{J_\alpha^\mathrm{DM}},
\end{equation*}
where $J_\alpha^\mathrm{DM}$ is the spatially-dependent Ly$\alpha$ intensity induced by the exotic energy injection. $S_\alpha$ depends on the present kinetic temperature $T_k$ but is otherwise independent of the energy injection process. Here we treat the Ly$\alpha$ deposition with the low energy photon module of version 1.1 of \dhis \cite{Liu:2019bbm, Sun:2022djj}.  We note that \dhis has since been updated with a more detailed treatment of low-energy photons and electrons that predicts the Ly$\alpha$ spectrum with significantly higher accuracy by tracking atomic hydrogen levels beyond the ground state~\cite{Liu:2023fgu}. However, due to the significantly prolonged run time associated with this improvement, we find it impractical to include it, leaving a more careful study to future work.

\subsection{Neutral Hydrogen Evolution} \label{sec:NeutralHydrogenEvolution}
\cmfast accounts for two effects which drive ionization of the IGM: weakly ionizing \textit{X}-ray emission and strongly ionizing UV emission.  The effect of weakly ionizing \textit{X}-ray emission is fully accounted for within the evolution of $x_e$, while the role of ionizing UV emission is calculated using an excursion-set approach~\cite{Mesinger:2007pd}.  Schematically, the density field is filtered on different radii, and if the expected ionizing radiation on any radius overcomes the number of recombinations, the pixel is considered ionized~\cite{Park:2018ljd}. In \cmfast, this procedure calculates a neutral hydrogen fraction $x_\mathrm{HI}^\mathrm{filter}$, which is then combined with $x_e$ to obtain the total neutral fraction
\begin{equation}
    x_\mathrm{HI} = \mathrm{max}[0, x_\mathrm{HI}^\mathrm{filter} - x_e].
\end{equation}
Note that the total $x_\mathrm{HI}$ is not used to update $x_e$. Since we have fully included the effects of exotic energy emission in the evolution of $x_e$ specified by~\eqref{eq:dxedzp-dTdzp}, we will then realize a correct evaluation of $x_\mathrm{HI}$.

\subsection{Summary of \cmfast Evaluation}
For a full review of \cmfast and its calculation of the 21-cm power spectrum, we refer readers to the most recent associated code paper and modeling procedures \cite{Park:2018ljd, Murray:2020trn, Qin:2020xyh, Munoz:2021psm}. Our interest is in how exotic energy injection modifies the 21-cm brightness temperature by contributing to the ionization of the IGM and the spin temperature evolution, which is evaluated using the \texttt{spin\_temperature} routine of \cmfast. However, a complete \cmfast evaluation, including the effects we attempt to model, requires additional inputs from routines that: \textit{i)} evolve the density and velocity fields (\texttt{perturb\_field}); \textit{ii)} compute the neutral hydrogen fraction including the effects of UV emission (\texttt{ionize\_box}), and \textit{iii)} calculate the brightness temperature from the density field, the velocity field, the ionization field, and spin temperature (\texttt{brightness\_temperature}).  The work we present here will only make modifications and introduce new functionality through  \texttt{spin\_temperature}, with the remaining routines left unmodified. However, we mention them here for completeness as they generate key inputs for our modeling procedure.

\section{Modeling Exotic Energy Injection in the 21-cm Power Spectrum}
\label{sec:Model}

Our code \dmcm provides detailed treatment of the spatially inhomogeneous injection and deposition of exotic energy through DM interactions. In tandem with standard astrophysical processes, this exotic energy injection and deposition will then determine the spatial and spectral morphology of the 21-cm emission. While \dmcm can be used for any source of energy injection,
in this work, we specifically consider exotic energy sourced by DM decays. Naturally, the volume and resolution of the simulations we intend to perform considerably guides the construction of our modeling procedure; all simulations in this work will be performed within a box of comoving size $(256 \, \mathrm{Mpc})^3$ at a comoving lattice resolution of $2\, \mathrm{Mpc}$. By comparison, we will advance our simulations in time using a timestep $\Delta z$ such that $\Delta z/(1+z) = 0.002$, corresponding to a comoving light-travel distance of approximately $ 1\,\mathrm{Mpc}$ at $z = 50$ and $2.5\,\mathrm{Mpc}$ at $z = 5$.
This is substantially finer than the default \cmfast timestep of $\Delta z/(1+z) = 0.02$.

DM may decay through a variety of channels, but over the astrophysical timescales relevant for modeling the 21-cm power spectrum, decay to any Standard Model particles will promptly generate photons, electrons/positrons, and neutrinos. We neglect the production of protons/antiprotons and atomic nuclei/antinuclei, which are typically subdominant, and we do not model the weak interactions of neutrinos with the Standard Model and their influence on cosmology. As a result, we need only model the role of photons and electrons/positrons injected by DM decay in determining the 21-cm signal. To do so, we will make use of \cmfast as a foundation while using \dhis to calculate the energy deposition to each of the channels described in Sec.~\ref{sec:Basics} for arbitrary spectra of injected electrons or photons. Going forward, we will use  ``electrons'' to also refer to positrons, unless explicitly stated otherwise.

\begin{figure}[!t]
    \begin{center}
    \includegraphics[width=0.49\textwidth]{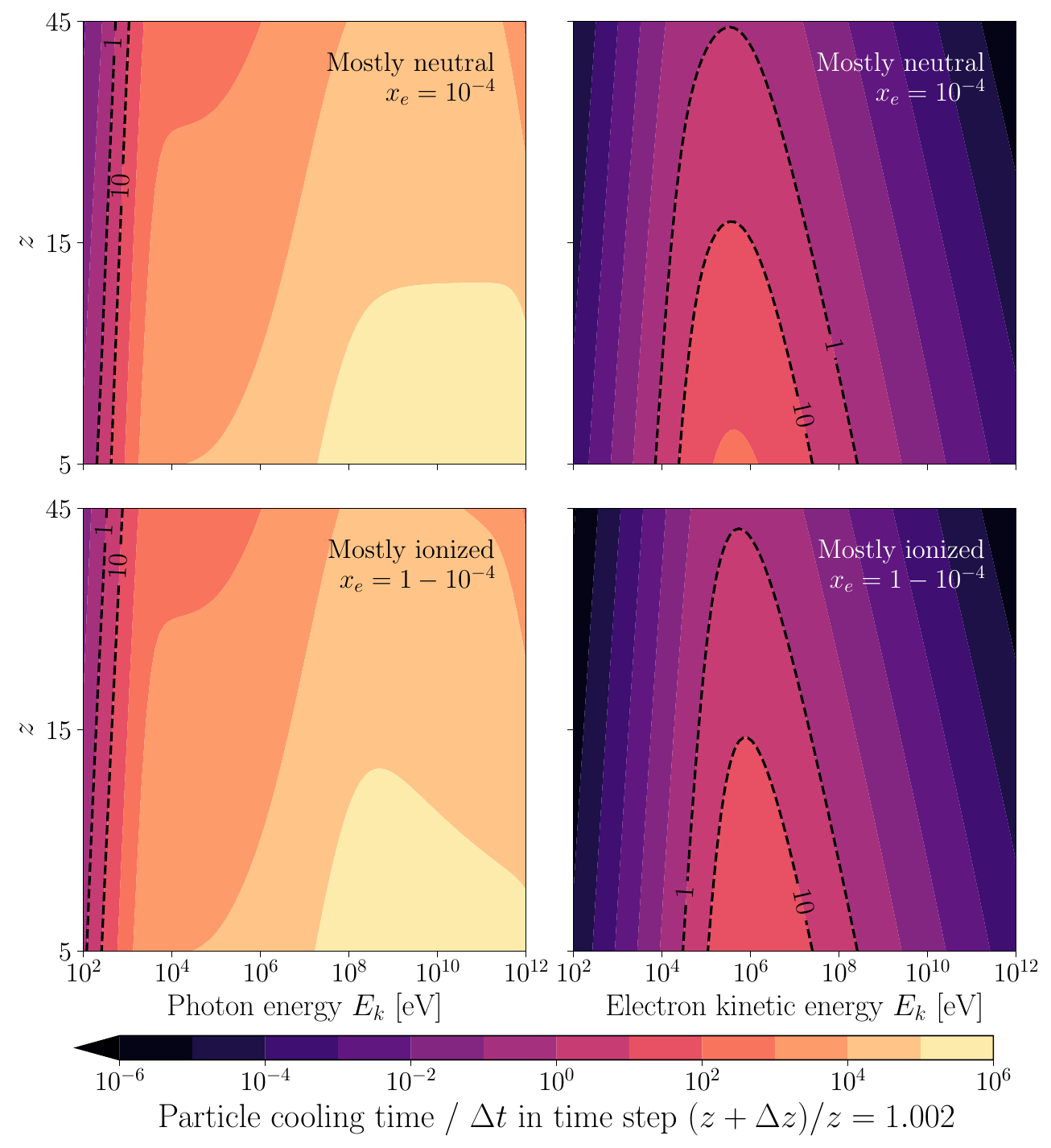}
    \vspace{-0.6cm}
    \caption{\textbf{Transparency window of photons and electrons during reionization.} The left two panels show that the universe is transparent to photons above $\sim$~keV during reionization, which lose little energy over a time step in our simulation of $\Delta z/(1+z)=0.002$.
    The right two panels show that for the majority of kinetic energies and redshifts, electrons lose most of their energy within a time step. We track long-lived photons in our simulation but assume all electrons deposit their energy within a time step.}
    \label{fig:transparency}
    \end{center}
\end{figure}

Before considering the construction of our calculation with \dmcm in detail, it is useful to examine the general characteristics of energy deposition via photons and electrons. In particular, the highly energy-dependent efficiency of energy deposition by photons and electrons strongly informs our modeling procedure. In Fig.~\ref{fig:transparency}, we study the efficiency of energy deposition (summed over all possible deposition channels) into mostly neutral and mostly ionized gas by examining the kinetic energy loss timescale $|d \log E_k / dt|^{-1}$ for photons and electrons calculated in \dhis. Then, as a joint function of kinetic energy and redshift, we examine the ratio of the kinetic energy loss timescale with the duration of the redshift timestep $\Delta z$ used in our simulations. 

\begin{figure*}[!ht]  
    \hspace{0pt}
    \vspace{-0.2in}
    \begin{center}
    \includegraphics[width=0.99\textwidth]{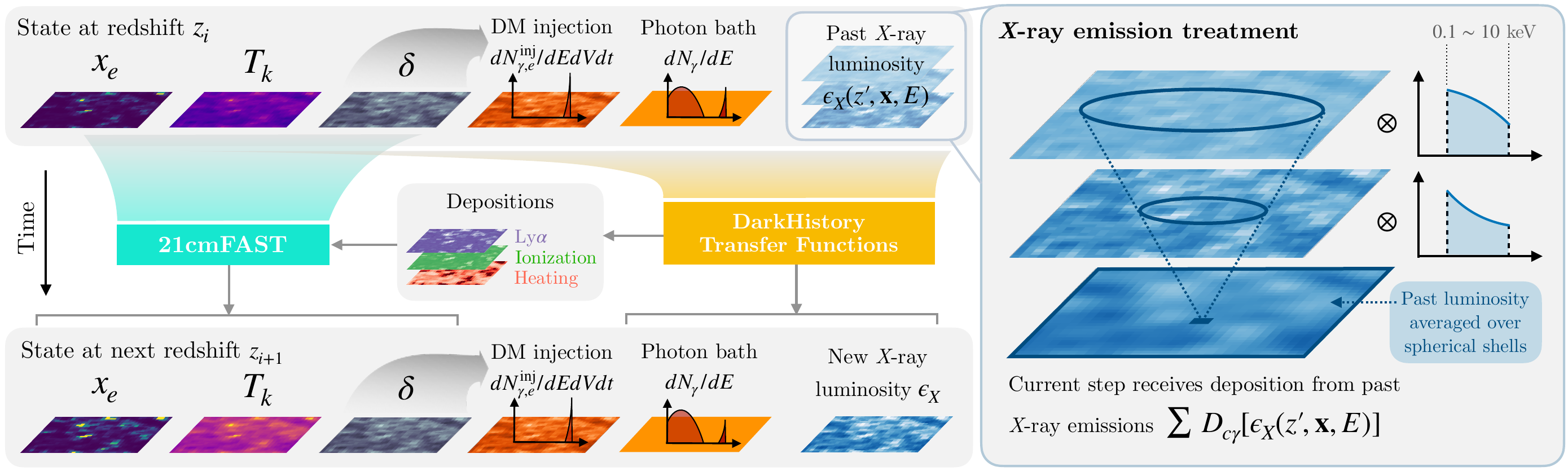}
    \caption{\textbf{A diagrammatic representation of the \dmcm redshift stepping process.} A single state of the simulation at $z$ consists of fields of density contrast $\delta$, the ionization fraction $x_e$, and kinetic temperature $T_k$,  the homogeneous, high-energy photon bath spectrum $dN^\text{bath}_\gamma/dE$, and the \textit{X}-ray emission history $\epsilon_X(z', \vx, E)$ cached over all prior states. The spatially inhomogeneous incident photon flux is determined from the uniform photon bath and the lightcone integral over the \textit{X}-ray emission history. Using \dhis transfer functions evaluated as a function of the local $\delta$, $x_\text{HI}$, and incident photon flux, the rate of energy deposition into the Ly$\alpha$, ionization, and heating channels is independently calculated for each location in the simulation volume. The energy deposition fields are then incorporated within a modified \cmfast step advancing the simulation state from $z_i$ to $z_{i+1}$. The \dhis transfer functions are also used to generate a new uniform high-energy photon bath and cache a new \textit{X}-ray relative luminosity field associated with production over the interval $z_i$ to $z_{i+1}$. In practice, the energy deposition procedure is subcycled with respect to the \cmfast timestep, see Sec.~\ref{sec:subcyling} for details. }
    \label{fig:Flow}
    \end{center}
    \vspace{-0.3cm}
\end{figure*}

In the case of electrons, with the exception of a modest transparency window at kinetic energies between $10^{5}\,\mathrm{eV}$ and $10^{7}\,\mathrm{eV}$, electrons deposit the majority of their energy within a single time step. Even ultrarelativistic electrons traveling along straight-line paths will deposit the majority of the energy before they travel the length of a lattice site, making their energy deposition instantaneous and on-the-spot to very good approximation. Even for electrons that have a cooling time of several time steps, intergalactic magnetic fields can potentially confine electrons to propagation distance which are much shorter than the lattice scale of our simulation. A magnetic field as weak as $10^{-20}\,\mathrm{G}$, for example, can confine electrons to a proper propagation length of less than $\sim 0.05\,\mathrm{Mpc}$, in comparison to our spatial resolution of $2\,\mathrm{Mpc}$. By comparison, lower limits on the intergalactic magnetic field based on gamma-ray observations of distant blazars are on the order of $B \gtrsim 10^{-16.5} \mathrm{ G}$ (see e.g.\ Ref.~\cite{AlvesBatista:2021sln} for a review). As a result, we expect even these electrons to deposit their energy in a highly on-the-spot manner. Moreover, since the cooling times for electrons are still short compared to the cosmological timescales over which we perform our simulation, treating their energy deposition as instantaneous remains a good approximation. 

Photons, however, have a more interesting behavior.
High-energy photons, which we define as photons at energies above $10\,\mathrm{keV}$, may travel for thousands or more timesteps before fully depositing their energy. This corresponds to comoving cooling distances on the scale of Gpc or greater, considerably larger than our $256\,\mathrm{Mpc}$ simulation boxes. As a result, while high-energy photons may be injected in a spatially inhomogeneous manner, the incident spectrum of high-energy photons at any one point receives contributions from all emission within the very large cooling distance that greatly exceeds the size of our simulations. This effectively averages out spatial inhomogeneities in the emission of high-energy photons, so we model them as a spatially homogeneous bath (evolving as photons are sourced by DM processes and scatter off the IGM).

On the other hand, at lower energies, the photon cooling distance can become small compared to the size of our simulation box, so photons maintain their inhomogeneities when averaged over a cooling distance. A similar challenge is encountered by \cmfast in the modeling of standard stellar \textit{X}-ray emission. Along the lines of \cmfast, we define \textit{X}-ray photons as those with energies between $100\,\mathrm{eV}$ and $10\,\mathrm{keV}$. We track the spatial dependence of their injection over the course of the simulation, and the energy-dependent manner in which they deposit their energy and attenuate beginning at their time of emission, in order to accurately model their effect on the brightness temperature. Though we implement our own treatment of photons in this energy range, it will in many ways parallel and improve upon that of \cmfast. 
Also like \cmfast\ we average over the line-of-sight angle for redshift space distortions.
The observed modes outside the ``wedge" are largely line-of-sight~\cite{Datta:2011hv,Parsons:2011ew}, though, and may have slightly higher power~\cite{HERA:2021noe}.

Between 10.2~eV and 100~eV, photons are characterized by a cooling distance smaller than the lattice resolution of our simulations and so require only an on-the-spot treatment analogous to electrons. We do not treat photons below the Ly$\alpha$ transition energy of 10.2~eV in this work.

With this multifaceted treatment of electrons and photons in mind, in Fig.~\ref{fig:Flow}, we provide a diagrammatic explanation of the \dmcm calculation which we detail in this section. The outline of our procedure is as follows:
\begin{enumerate}
    \item Following the structure of \cmfast, we initialize the overdensity field $\delta$, two-phase ionization $x_e$ and $x_\text{HI}$, kinetic temperature $T_k$ and spin temperature $T_S$  at $z=45$. For a given DM model, we add or subtract a spatially homogeneous contribution from the $T_k$ and $x_e$ fields so that their global averages match the values predicted by \dhis.\footnote{For relatively rapid decay scenarios, a modification of the calculation of adiabatic index relating $\delta T_k$ to $\delta$ that accounts for the effect of DM heating may be necessary \cite{Munoz:2023kkg}. This treatment would also need to be extended to modify initial conditions for $x_e$.} We also initialize a homogeneous bath spectrum of photons $dN_\gamma/dE$ produced by DM processes which occur at $z > 45$ as predicted by \dhis. 
    \item At each time, to evolve $x_e$, $T_k$ and $T_S$ a single discretized step, in addition to contributions coming from star formation already included in \cmfast, we need the additional DM terms $\textcolor{red}{dT_k^\text{DM} / dz}$, $\textcolor{red}{dx_e^\text{DM} / dz}$ and $\textcolor{red}{J_\alpha^\text{DM}}$, corresponding to energy deposition into heating, ionization and Ly$\alpha$ photons. There are three main energy deposition contributions to these terms that we have to treat separately:
    \begin{enumerate} 
        \item DM processes occurring within each cell, which deposit their energy promptly and locally. This includes energy injected in the form of electrons/positrons and photons with energy between 10.2~eV and $100 \,\mathrm{eV}$;
        \item \textit{X}-ray photons (defined as photons with energies between $100\,\mathrm{eV}$ and $10\,\mathrm{keV}$ at the time of emission) that have an energy-loss path length comparable to the size of the simulation arrive at each cell from neighboring cells. The spectrum incident on each cell is different, and has to be computed by integration along the lightcone, accounting for redshifting and attenuation; and
        \item High-energy ($> 10 \,\mathrm{keV}$) photons with energy-loss path lengths much larger than both the size of the simulation and the  Hubble length. We treat these photons with a single homogeneous high-energy photon spectrum.  
    \end{enumerate}
    \item After calculating the energy deposition terms, we also need to obtain:
    \begin{enumerate}
        \item The spectrum $dN_X/dE$ and spatial distribution of \textit{X}-rays emitted during the step. This information is cached for future use in the lightcone integration of \textit{X}-ray photons of subsequent steps.
        \item The change to the homogeneous photon spectrum $dN^\mathrm{bath}_\gamma/dE$ due to interactions in all cells, which is then passed to the next step.
    \end{enumerate}
\end{enumerate}
In the following subsections, we will detail the development of transfer functions from \dhis and their incorporation within \dmcm that map input photon and electron spectra into energy deposition channels and secondary photon production, enabling this modeling procedure. These transfer functions are made publicly available for use with \dmcm in \cite{foster_2023_10397814}. 
In Apps.~\ref{app:AdiabaticEvolution}, \ref{app:SFRD}, and \ref{app:DM}, we perform extensive convergence tests between our procedure here and both \cmfast and \dhis, finding excellent agreement and consistency. 

\subsection{Exotic Energy Deposition from Photons and Electrons} 
\label{sec:DMInj}
Given any DM energy injection process, we now want to determine the terms \textcolor{red}{$d x_e^\mathrm{DM} / dz$}, \textcolor{red}{$d T_k^\mathrm{DM}/dz$} in~\eqref{Eq:ODE}, as well as \textcolor{red}{$J_\alpha^\text{DM}$} for the determination of $x_\alpha$ in~\eqref{Eq:x_alpha}. As reviewed in Sec.~\ref{sec:Basics}, to model 21-cm cosmology in the presence of exotic energy injection from DM, we must be able to calculate how particles generated through DM processes across a wide range of energies deposit energy into heating, ionization, and Ly$\alpha$ excitation through scatterings off the IGM and the radiation field.\footnote{We consider only the contribution of the CMB to the background radiation field, though the early starlight background may also contribute. We leave a more detailed treatment of the stellar light background to future work.} These processes are intrinsically sensitive to the local state of the IGM, \textit{i.e.}, the baryon density and ionization fraction. The same scattering events which deposit energy will also deplete DM byproducts over time while potentially generating new secondary photons. It is thus critical that we model the time-evolving abundance and spectral energy distribution of photons generated by exotic energy injection processes in order to provide accurate input spectra for energy deposition calculations at each time in the simulation.

We perform the modeling of energy deposition and secondary photon produced in scattering events with \dhis, a code designed to describe the cascade of particle production and energy deposition from exotic processes such as DM decay at times before recombination until the end of reionization. In particular, \dhis evolves the spectrum of photons as well as the matter temperature and ionization levels of hydrogen and helium assuming a homogeneous universe at mean baryonic density. \dhis is able to accelerate this computation by pre-computing the total output (in terms of deposition and secondaries) of the above processes over a range of injection energies, ionization fractions of hydrogen and helium, and redshifts, compiling the results into transfer functions, and then interpolating over them in an actual evolution \cite{Liu:2019bbm, Sun:2022djj}.

We build on this treatment to generate transfer functions that act on input spectra of photons and electrons/positrons, parametrized in terms of the local density and ionization fraction, to determine how each particle cools. Specifically, these transfer functions will act on an input spectrum $dN^\text{in}_{i}/dE$ defined in terms of the number density spectrum $dN^\text{in}_i/ dE dV$ and the average baryon number $\bar n_b$ as 
\begin{equation}
    \frac{dN^\text{in}_i}{dE} = \frac{1}{\bar n_b} \frac{dN^\text{in}_i}{dE dV},
\end{equation}
where $i = \gamma$ for photons and $i = e$ for electrons/positrons. These transfer functions will be evaluated as a function of baryonic overdensity through the local density contrast $\delta$ and the total local ionization fraction through $1-x_\mathrm{HI}$. Since our transfer functions depend on the total local ionization fraction, we must use $1-x_\mathrm{HI}$ rather than $x_e$, which accounts only for \textit{X}-ray ionization.

\subsubsection{Electron Transfer Functions}
We first consider the construction of the relevant transfer functions for the relatively simpler case of electrons. As we have argued, electrons deposit their energy in a manner that is, to good approximation, instantaneous and on the spot. This enables us to accurately describe electron processes with just two transfer functions: $D_{ce}$, which maps an input spectrum of electrons into the energy they deposit into heat ($\Delta T_k$), ionization ($\Delta x_e)$, and Ly$\alpha$ excitation ($J_\alpha$) occurring over a redshift duration $\Delta z$, and  $T_{\gamma e}$, which maps an input spectrum of electrons into an outgoing spectrum of photons for each cell. Schematically, these transfer functions are applied as 
\begin{equation} \label{eq:electronTF}
\begin{gathered}
    \frac{dN_\gamma^\text{out}}{dE} = T_{\gamma e}(\delta, x_\mathrm{HI} | \Delta z) \frac{dN_e^\text{in}}{dE} \\
    \bbm \Delta T_k \\ \Delta x_e \\ J_\alpha \ebm = D_{ce}(\delta, x_\mathrm{HI} | \Delta z) \frac{dN_e^\text{in}}{dE},
\end{gathered}
\end{equation}
where we have made explicit the dependence of these transfer functions on the baryon density through the density contrast $\delta$, the ionization fraction through $x_\mathrm{HI}$, and the duration of interval $\Delta z$. Note that although exotic energy deposition to ionization is assigned to $x_e$, the transfer functions depend on $x_\mathrm{HI}$, the neutral fraction accounting for \textit{X}-rays, UV radiation and DM processes. The subscripts $\gamma$, $e$, and $c$ denote input/output to photons, electrons, and deposition into channel $c$ (indexing over heat, ionization, and Ly$\alpha$ excitation) respectively. The output spectrum is defined with the same convention relative to the average baryon number density. The transfer functions are constructed by applying the inverse Compton scattering (ICS) and positron procedures in \dhis to cool/annihilate high energy electrons/positrons. One output of these procedures is a spectrum of low-energy electrons; we fully deposit the available energy into secondary photons and the various deposition channels by interpolating older results from MEDEA for the behavior of these low-energy electrons~\cite{Evoli:2010zz, evoli2012energy}, following the standard method in \dhis v1.1 \cite{Liu:2019bbm}. Since there will be no outgoing electron spectrum, we have no need for a transfer function $T_{ee}$.

\subsubsection{Photon Transfer Functions}

Just like for electrons, we will generate a transfer function $D_{c \gamma}$ relating an input spectrum of photons per baryon  $dN^\mathrm{in}_{\gamma}/dE$ to energy deposition into the three relevant channels. However, rather than generating a single transfer function that maps an input spectrum of photons to an output spectrum of photons, we will consider two transfer functions: $P_{\gamma \gamma}$ and $T_{\gamma \gamma}$. The transfer function $P_{\gamma \gamma}$ is the ``propagating photon transfer function", which maps an input photon spectrum $dN_{\gamma}^\mathrm{in}/dE$ into an output spectrum of propagating photons $dN_{\gamma}^\mathrm{out, prop.}$ that did not undergo a scattering during the redshift interval $\Delta z$. By contrast, $T_{\gamma\gamma}$ is the ``scattered photon transfer function" which maps an input photon spectrum $dN_{\gamma}^\mathrm{in}/dE$ to the spectrum of outgoing photons from scattering events. Collectively, we have
\begin{equation} \label{eq:photonTF}
\begin{gathered}
    \frac{dN_\gamma^\text{out}}{dE} = \big[ P_{\gamma \gamma}(\delta, x_\mathrm{HI} | \Delta z) + T_{\gamma \gamma}(\delta, x_\mathrm{HI} | \Delta z) \big]\frac{dN_\gamma^\mathrm{in}}{dE} \\
    \bbm \Delta T_k \\ \Delta x_e \\ J_\alpha \ebm = D_{c\gamma }(\delta, x_\mathrm{HI} | \Delta z) \frac{dN_\gamma^\text{in}}{dE}.
\end{gathered}
\end{equation}
The decomposition of the photon-to-photon transfer function into a propagating and scattered part enables a more detailed treatment of the direction photons travel and the locations at which they deposit their energy. In particular, photons from the propagating transfer function travel unperturbed along their trajectories, while we assume the trajectories of scattered photons are isotropized. We will take advantage of this modeling flexibility in Sec.~\ref{sec:Implementation} to develop a detailed treatment of \textit{X}-ray photons that travel moderate lengths before fully depositing their energy. 

\subsubsection{Numerical Implementation and Interpolation Table Construction}

To build the relevant transfer functions, we modify \dhis and its data files to calculate interaction rates for photon and electron interactions as a function of both baryon density and ionization fraction. We discretize the photon and electron kinetic energy spectra into 500 log-spaced bins spanning $10^{-4}$ to $10^{12}$ eV, and choose $10$ log-spaced redshifts between $z = 5$ and $z = 50$. We also select a timestep size $\Delta z$, which is fiducially taken to satisfy $\Delta z/(1+z)=0.002$. At each of our $10$ redshifts, we generate a discretized transfer function matrix evaluated at baryon overdensity $\delta$ and ionization fraction $x_e$ by injecting monochromatic photon input spectra and evaluating the resulting output spectra and energy deposition into each channel over a single timestep of duration $\Delta z$ using \dhis. Photon transfer function matrices are generated for 10 values of neutral fraction $x_\mathrm{HI}$
between $10^{-5}$ and $1-10^{-5}$ and 10 values of the baryonic overdensity $\delta$ between $10^{-3}$ and $10$. In total, this provides a grid of transfer function matrices of size (10, 10, 10) in $(z, \delta, x_\mathrm{HI})$; to evaluate transfer function matrices at values of $(z, \delta, x_\mathrm{HI})$ between the evaluation points, we interpolate. We follow a similar procedure, injecting monochromatic electron spectra to develop electron transfer function matrices, though we evaluate at 30 points in $x_\mathrm{HI}$ for a better interpolation resolution. At present, the transfer function table sizes are limited by the available GPU memory as placing the tables into GPU is crucial for evaluation speed. This limitation on the interpolation table resolution leads to a small but finite interpolation error which we study in App.~\ref{app:DM}. We note that the memory footprint of the transfer functions can be reduced by more than $\mathcal{O}(100)$ by replacing them with dense neural networks, similar to transfer functions in v1.1 of \dhis \cite{Sun:2022djj}. This would also enable the possibility of more detailed modeling requiring additional parameters, such as an independent singly ionized helium fraction $x_\text{HeII}$. We leave the implementation to future work.

We caution prospective users of the \dmcm code framework that we have framed this discussion in terms of an input spectrum of photons/electrons $dN_{\gamma/e}/dE$ for the sake of clarity. Internally, our discretized transfer function matrices operate on the vector $\mathbf{N}$, with elements $\mathbf{N}_i$ defined as the number density of particles with kinetic energy between bin edges $E_i$ and $E_{i+1}$ divided by the global average baryon density, matching the convention in \dhis.

\subsection{Exotic Energy Injection in \dmcm} 
\label{sec:Implementation}

Equipped with the transfer functions developed in Sec.~\ref{sec:DMInj}, we are now prepared to develop our full treatment of the energy deposition through prompt processes, \textit{X}-rays, and high-energy photons. A full treatment of all these processes includes the development of a custom caching and lightcone integration procedure; for computational efficiency, we also design an efficient subcycling scheme that decouples the \cmfast time steps from the time steps used for our custom energy injection scheme, which requires high time resolution, to maintain good spatial resolution.

\subsubsection{Prompt Injection}
\label{sec:OTS}
The simplest procedure by which exotic energy injection is realized in \dmcm is through ``on-the-spot" processes, characterized by interactions that cause energetic particles to deposit their energy before they can propagate on appreciable length scales. We begin at redshift $z_{i}$, when we have evaluated the density field $\delta(z_{i}, \mathbf{x})$ with \cmfast. The spatially inhomogeneous rate of DM depletion events, \textit{i.e.}, decays or annihilation, can be evaluated in terms of the local density $\delta$ and the dark matter parameters, such as mass $m_\chi$ and its lifetime $\tau$ or velocity-averaged cross-section $\langle \sigma v \rangle$. For DM decay, the rate of injection events per unit volume into decay products is given by
\begin{equation}
    \frac{dN^\text{inj}}{dVdt}(\delta, m_\chi)=\bar{\rho}_\text{DM}(1+\delta)/(\tau m_\chi),
\end{equation}
where $\bar \rho_\text{DM}$ is the mean physical DM density at the time of injection. For a given decay channel, we can calculate the spectrum of secondary photons and electrons per injection event using \texttt{PPPC4DMID} \cite{Cirelli:2010xx}, which we denote $dN_\gamma/dEdN^\text{inj}$ and $dN_{e}/dEdN^\text{inj}$, respectively. In this work, we consider only monochromatic decays to either photons or electrons as illustrative example cases, though more general spectra are trivially accommodated in \dmcm. From these quantities, we can calculate a differential number spectrum of injected photons or electrons as
\begin{equation}
    \frac{dN^\mathrm{ots}_{\gamma/e}}{dE} = \frac{\Delta t(z_i, z_{i+1})}{\bar n_b}\frac{dN^\mathrm{inj}}{dVdt}\cdot\frac{dN_{\gamma/e}}{dEdN^\text{inj}},
\end{equation}
where $\Delta t(z_i, z_{i+1})$ is the time between the start of our step at $z_i$ and the end of our step at $z_{i+1}$. This allows us to use our \dmcm transfer function to calculate $dT_k/dz$, $dx_e/dz$, and $J_\alpha$ in a step as
 \begin{equation}
 \begin{split}
    \begin{bmatrix}
           \Delta T_k (z_i, \mathbf{x}) \\
           \Delta x_e (z_i, \mathbf{x}) \\
           J_\alpha(\mathbf{x})
    \end{bmatrix}_\mathrm{ots}\hspace{-0.4cm} &= D_{c\gamma}(z_i, \mathbf{x}) \frac{dN^\mathrm{ots}_{\gamma}}{dE}(z_i, \mathbf{x}, E) \\
    &\phantom{=}+  D_{ce}(z_i, \mathbf{x})\frac{dN^\mathrm{ots}_{e}}{dE}(z_i, \mathbf{x}, E),
  \end{split}
\end{equation}
where the transfer functions inherit spatial dependence through their dependence on $\delta$ and $x_\mathrm{HI}$, while the injected spectrum normalizations are spatially dependent only through their dependence on $\delta$.

Similarly, a spatially dependent outgoing photon spectrum $dN_{\gamma}^\mathrm{ots, out}/dE$ is calculated by
\begin{equation}
\begin{split}
\frac{dN_{\gamma}^\mathrm{ots, out}}{dE}(\mathbf{x}, E) &=  T_{\gamma\gamma}(z_i, \mathbf{x}) \frac{dN^\mathrm{ots}_{\gamma}}{dE}(z_i, \mathbf{x}, E) \\
&\phantom{=} + T_{\gamma e}(z_i, \mathbf{x})\frac{dN^\mathrm{ots}_{e}}{dE}(z_i, \mathbf{x},E).
\end{split}
\end{equation}
These are secondary photons that are produced by the same scatterings of energetic photons and electrons that produce heat, ionization, and Ly$\alpha$ excitation. Photons below $10\,\mathrm{keV}$ will be accounted for within the \textit{X}-ray treatment while those above $10\,\mathrm{keV}$ will be accounted for within the high-energy photon treatment. Note that we assume that energetic electrons promptly deposit their energy and are converted to photons. We will return to the fate of the outgoing photon spectrum in Sec.~\ref{sec:Caching}.

\subsubsection{Homogeneous Bath Injection}
\label{sec:Bath}
High-energy photons ($E>10\,\mathrm{keV}$) propagate on cosmological scales comparable to the Hubble horizon before they deposit their energy, meaning that the energetic photon distribution is well-approximated as spatially homogeneous. For these photons, we need only track and evolve a single photon spectrum $dN^\mathrm{bath}_{\gamma}/dE$ over the course of the simulation. \dhis evolves such a spectrum, which initializes our high-energy homogeneous photon bath at the beginning of our \dmcm simulations. After this time, the bath is self-consistently evolved, which we describe at further length in Sec.~\ref{sec:Caching}. By convention, we assume photons with energies above $10\,\mathrm{keV}$ are well-described as spatially homogeneous and so live within our bath. This is consistent with \cmfast, which supports an \xray spectrum going up to 10 keV. 

From the bath spectrum $dN^\mathrm{bath}_{\gamma}/dE$ evaluated at time $z_i$ , we evaluate
\begin{equation}
 \begin{split}
    \begin{bmatrix}
           \Delta T_k (z_i, \mathbf{x}) \\
           \Delta x_e (z_i, \mathbf{x}) \\
           J_\alpha(\mathbf{x})
    \end{bmatrix}_\mathrm{bath}\hspace{-0.6cm}&= D_{c\gamma}(z_i, \mathbf{x}) \frac{dN^\mathrm{bath}_{\gamma}}{dE}(z_i, E).
    \end{split}
\end{equation}
Like in the case of on-the-spot injection, we also compute a spatially dependent outgoing secondary photon spectrum
\begin{equation}
\begin{split}
\frac{dN_{\gamma}^\mathrm{bath, out}}{dE}(z_i, \mathbf{x}, E) &=  T_{\gamma\gamma}(z_i, \mathbf{x}) \frac{dN^\mathrm{bath}_{\gamma}}{dE}(z_i, E).
\end{split}
\end{equation}
Note that although the incoming bath spectrum is spatially homogeneous, the energy deposition and the outgoing photon spectrum at each cell depend on the local overdensity and ionization fraction, and is spatially inhomogeneous. The dependence of the transfer functions on $\mathbf{x}$ serves as a reminder of this fact.

\subsubsection{\textit{X}-Ray Spectrum Injection}
\label{sec:XRayDeposition}
Like in \cmfast, the treatment of \textit{X}-ray photons ($100\,\mathrm{eV} < E < 10\,\mathrm{keV}$) is the most challenging and involved part of our energy deposition procedure, as they propagate on observationally relevant Mpc-scales. Photons produced with energies below 100~eV have very short propagation lengths and are already accounted for in the on-the-spot deposition described in the previous section. Photons above 10~keV have such long propagation lengths that they are accurately described by our homogeneous bath described in Sec.~\ref{sec:Bath}.

To determine the \xray spectrum incident on a particular cell, due to the intermediate nature of the path length, we need to integrate the contribution from all cells along the lightcone of the cell of interest. To do this in full, we would need to save the \xray spectrum of every cell for all redshifts prior to the current one even under our isotropized \xray emission assumption, which is computationally intractable. Instead, we make the simplifying assumption that at the current redshift $z_i$, the spectrum of photons emitted from $\vx$ at redshift $z_e$ can be written as 
\begin{equation} \label{eq:xrayseparabliblity}
    \frac{dN_X}{dE d\tau}(z_i, \vx, E | z_e)\approx\frac{dN_X}{dE d\tau}(z_i, E |z_e)\,\tilde\epsilon_X(\vx | z_e).
\end{equation}
Here, $dN_X/d\tau dE(z_i, E |z_e)$ is a photon emission spectrum rate with respect to conformal time which has been consistently attenuated and redshifted from emission at $z_e$ to $z_i$, and $\tilde\epsilon_X(\vx| z_e)$ is the spatially inhomogeneous \textit{relative} luminosity of \textit{X}-rays at the redshift of emission $z_e$ that averages to 1. In other words, we assume that the emitted spectrum at every point at $z_e$ in the simulation can be characterized by a universal spectral shape, differing point-by-point only by a normalization described by $\tilde\epsilon_X$. Precise details of this attenuation and redshifting will be provided in Sec.~\ref{sec:Caching}.

Subject to this assumption, the number spectrum of previously emitted photons at spatial location $\vx$ at redshift $z_i$ can be evaluated with the discretized lightcone integral summing over the $\epsilon_X$ which were evaluated and cached at previous redshift steps by 
\begin{equation}
\begin{split}
\label{eq:DiscreteLightCone}
&\frac{dN_X^\mathrm{lightcone}}{dE}(z_i,\vx,E) = \\
&\hspace{0.2cm}\sum_{j} \int \frac{d^2\hat n}{4\pi} \tilde\epsilon_X(\mathbf{x} - \hat nR(z_{j}, z_i) |z_{j}) \frac{dN_X}{dE}(z_i , E | z_j),
\end{split}
\end{equation}
where $R(z_1, z_2)$ is the comoving distance traversed by a photon between redshifts $z_1$ and $z_2$, $dN_X/dE(z_i , E | z_j)$ is the redshifted and attenuated spectrum of total photons per baryon emitted between $z_{j}$ and $z_{j-1}$.
We approximate the surface integral in~\eqref{eq:DiscreteLightCone} by a spatial average so that we obtain
\begin{equation}
\begin{split}
\label{eq:FullyDiscreteLightCone}
&\frac{dN_X^\mathrm{lightcone}}{dE}(z_i,\vx,E) = \\
&\hspace{0.2cm}\sum_{j} \bar\epsilon_X(\mathbf{x}  |z_{j}, R(z_i, z_{j}), R(z_i, z_{j-1})) \frac{dN_X}{dE}(z_i , E | z_j),
\end{split}
\end{equation}
where $\bar\epsilon_X(\mathbf{x} |z_{j}, R_1, R_2)$ is the average value of $\tilde\epsilon_X$ on the annulus defined centered at $\mathbf{x}$ defined by radii $R_1$ and $R_2$. 

Like we did in Sec.~\ref{sec:OTS}, we can use this spectrum to calculate the energy deposition
\begin{equation}
 \begin{split}
    \begin{bmatrix}
           \Delta T_k (z_i, \mathbf{x}) \\
           \Delta x_e (z_i, \mathbf{x}) \\
           J_\alpha(\mathbf{x})
    \end{bmatrix}_X\hspace{-0.3cm} &= D_{c\gamma}(z_i, \mathbf{x}) \frac{dN_X^\mathrm{lightcone}}{dE}(z_i, \mathbf{x}, E)
    \end{split}
\end{equation}
and outgoing photon emission spectrum 
\begin{equation}
\begin{split}
\frac{dN_{X}^\mathrm{out}}{dE}(z_i, \mathbf{x}, E) &= T_{\gamma\gamma}(z_i, \mathbf{x}) \frac{dN_X^\mathrm{lightcone}}{dE}(z_i, \mathbf{x}, E)
\end{split}
\end{equation}
that is produced over the interval $z_{i}$ to $z_{i+1}$.

\subsubsection{Caching, Redshifting, and Attenuating}
\label{sec:Caching}
While we have fully described our energy deposition procedure, we must still describe how the homogeneous bath spectrum, \textit{X}-ray emission histories, and evolved \textit{X}-ray emission spectra are evaluated. As we have already described the manner in which outgoing photon spectra are evaluated, the homogeneous bath spectrum and \textit{X}-ray emission spectrum can be straightforwardly constructed.

In the case of the homogeneous bath spectrum, we must advance it from $z_i$ to $z_{i+1}$ by accounting for the attenuation it undergoes due to scatterings (which are the same scatterings that cause it to deposit energy), redshifting of energies, and new bath-energy photons that are sourced by DM processes. This update step takes the form
\begin{equation}
\begin{split}
    \frac{dN_\gamma^\mathrm{bath}}{dE}(z_{i+1}, E) &= P_{\gamma\gamma}(z_{i+1}, z_{i}) \frac{dN_\gamma^\mathrm{bath}}{dE}(z_i, E) \\
    &\phantom{=}+ \frac{dN_\gamma^\mathrm{bath, source}}{dE}(z_i, E),
\end{split}
\end{equation}
where the propagation transfer function $P_{\gamma\gamma}$ which advances the spectrum from $z_{i}$ to $z_{i+1}$ is evaluated using the global average hydrogen ionization fraction $1-\bar x_\mathrm{HI}$ as calculated by \cmfast at $z_i$ and accounts for attenuation and redshifting. The bath source is calculated as
\begin{equation}
\frac{dN_\gamma^\mathrm{bath, source}}{dE}  = \bigg\langle \frac{dN_{\gamma}^\mathrm{ots, out}}{dE} + \frac{dN_{\gamma}^\mathrm{bath, out}}{dE} \bigg\rangle \theta(E - 10~\text{keV}),
\end{equation}
where $\langle \, \rangle$ indicates a spatial average and we have thresholded above 10~keV, as those photons will contribute to cached \textit{X}-ray spectrum and emission box instead. Similarly, the outgoing \textit{X}-ray spectrum does not contribute here as it does not have support above 10~keV.

Next, now that we have computed the outgoing X-ray spectrum for each cell, we want to apply our simplifying assumption to reduce these spectra to a new universal \textit{X}-ray spectrum and a spatially inhomogeneous relative \textit{X}-ray luminosity box. First, we calculate the spatially varying total \textit{X}-ray emission from each location in the simulation by
\begin{equation}
\begin{split}
\frac{dN_X}{dE}&(z_{i+1}, \mathbf{x}, E | z_{i+1}) = \frac{dN_{X}^\mathrm{out}}{dE} \\ +
&\left[\frac{dN_{\gamma}^\mathrm{ots, out}}{dE} + \frac{dN_{\gamma}^\mathrm{bath, out}}{dE}\right]\theta( 10~\text{keV} - E)\,,
\end{split}
\end{equation}
which includes the spatially varying outgoing X-rays produced by incoming X-rays scattering within each cell and contributions from X-rays coming from both scattering of prompt X-rays from the DM process, and scattering of the homogeneous high-energy spectrum. No attenuation through the $P^{\gamma\gamma}$ transfer function is necessary as these photons have been produced in scattering events but have not yet scattered themselves. However, they must be appropriately redshifted to the end of the step. We then calculate the universal \textit{X}-ray spectrum by
\begin{equation}
\frac{dN_X}{dE}(z_{i+1}, E | z_{i+1}) = \bigg\langle \frac{dN_X}{dE}(z_{i+1}, \mathbf{x}, E | z_{i+1}) \bigg\rangle\,.
\end{equation}
Furthermore, we construct the relative luminosity box by 
\begin{equation}
\begin{split}
\tilde\epsilon_X(z_{i+1}, \mathbf{x}) = \frac{\int dE \frac{dN_{X}}{dE}(z_{i+1}, \mathbf{x}, E)}{\left\langle\int dE \frac{dN_{X}}{dE}(z_{i+1}, \mathbf{x}, E) \right\rangle},
\end{split}
\end{equation}
where both integrations cover $E$ from 0.1~keV to 10~keV. Note that this does not preserve the spectrum at each location in the simulation, but it does preserve the spatial variation in the total energy emitted in \textit{X}-rays. This is the best achievable result under the constraint of energy conservation in combination with our separability approximation in \eqref{eq:xrayseparabliblity}. We have checked and found that in regimes where either the DM-sourced prompt injection or the bath photons dominate, the spectral shape of \xray photons produced in each simulation cell is indeed very similar, with the $dN/dE$ values consistent with each other at the $<10$\% level after adjusting for normalization.

In much the same way that we updated the bath spectrum, we update the previously evaluated (but not the presently evaluated) \textit{X}-ray emission spectrum as 
\begin{equation}
    \frac{dN_X}{dE}(z_{i+1}, E | z_j)= P_{\gamma\gamma}(z_{i+1}, z_{i}) \frac{dN_X}{dE}(z_i, E| z_j),
\end{equation}
where $P_{\gamma\gamma}$ accounts for propagation, attenuation and redshifting of the spectrum. Note that all prior cached \textit{X}-ray spectra must be updated in this manner to enable an accurate lightcone calculation. Additionally, to reduce the number of terms included in the sum in~\eqref{eq:FullyDiscreteLightCone} that will be performed to advance the state from $z_{i+1}$ to $z_{i+2}$, we dump any \textit{X}-ray spectra associated with emission at $z_j$ into the bath spectrum if $R(z_{i+1}, z_j)$ is larger than the half length of the simulation box. This amounts to treating emission along the lightcone at comoving distances greater than the box volume as homogeneous while preserving the total photon energy. This is because smoothing on scales larger than the box radius will, to a good approximation, return only the average over the box and is therefore straightforwardly incorporated as a bath contribution. Note that although this procedure homogenizes, it correctly includes all attenuation effects through the consistent evolution of the \textit{X}-ray spectra up until this point through $P_{\gamma \gamma}$.

This concludes the set of steps that must be performed to keep our cached data used in energy deposition up-to-date. Iterating through this procedure shown in Fig.~\ref{fig:Flow}, we are then able to include arbitrary exotic energy injection into the \cmfast framework.

\subsubsection{Lightcone Integration with Subcycling}
\label{sec:subcyling}
The construction of~\eqref{eq:FullyDiscreteLightCone} directly relates the spatial resolution of the lightcone integral to the resolution of the timestepping performed in advancing $z_{j-1} \rightarrow z_{j} = z_{j-1} + \Delta z$. This differs from the default \textit{X}-ray treatment of \cmfast, but enables a self-consistent and globally energy conserving treatment of \textit{X}-ray emission spectra that experience attenuation via the optical depth and do not have a spectral morphology at the time of emission which is constant in $z$.

On the other hand, this requires a very small timestep $\Delta z$ in order to achieve good spatial resolution. In a standard \cmfast simulation, while the lattice resolutions is typically roughly $2\,\mathrm{Mpc}$, the comoving distance associated with a standard redshift step of $\Delta z/(1+z) = 0.02$ is roughly $25\,\mathrm{Mpc}$ at $z = 5$. Obtaining spatial resolution down to the lattice scale, which is necessary to accurately resolve the propagation of low energy \textit{X}-rays that are strongly attenuated, we must then use a timestep which is considerably smaller than \cmfast uses by default.

Operating with this extremely small step size would be computationally costly and require an impractical amount of disk space due to \cmfast's built-in caching mechanism. To address this challenge, we subcycle our energy deposition procedure using fine timesteps $\Delta z_\mathrm{fine}$ with a \cmfast update performed with the coarse timestep $\Delta z_\mathrm{coarse}$. For a subcycling ratio $N_\mathrm{sub}$, we perform $N_\text{sub}$ of our custom energy deposition steps, accumulating the total energy deposition into each channel ($x_e$, $T_k$, and $J_\alpha$) while advancing the bath and \textit{X}-ray spectra and caching the relative \textit{X}-ray brightness $\epsilon$ on each fine step. After $N$ fine steps have been performed, we perform a \cmfast step, with
\begin{equation}
    \Delta z_\text{coarse}/(1+z)=1.002^{N_\text{sub}}-1,
\end{equation}
injecting the total accumulated energy deposition.

All simulations in this work are performed with $\Delta z_\mathrm{fine}/(1+z) = 0.002$ and $N_\mathrm{sub} = 10$ so that $\Delta z_\mathrm{coarse}/(1+z)\approx0.02$, reproducing the recommended \cmfast timestep. In App.~\ref{app:SpatioTemporal}, we systematically vary the choice of $\Delta z_\mathrm{fine}$, finding convergence to within subpercent accuracy for our fiducial choice.

\subsection{Comparison with \cmfast} \label{sec:XrayComparison}

In App.~\ref{app:FullTriangle}, we perform a cross-check of this framework by reproducing the \textit{X}-ray treatment of \cmfast with \dmcm. Here, we provide a more general overview, comparing the differences between the two codes. However, we emphasize that in general, the framework of \dmcm does not replace any functionality or modeling of \cmfast and instead allows for a user-defined injection of photons or electrons with custom spatial and spectral morphology, making it a highly flexible tool for both studies of DM and beyond.

In the most general sense, our \textit{X}-ray treatment and that of \cmfast are highly similar, with \cmfast performing a lightcone integration analogous to~\eqref{eq:DiscreteLightCone}. However, in \cmfast, the surface-averaged emission is calculated using an extended Press-Schechter treatment to calculate a halo mass function that is parametrically related to a SFRD and associated \textit{X}-ray emission spectrum. This extended Press-Schechter calculation is performed by backscaling the present-time density field and is independently evaluated at each redshift step. By contrast, we cache our emission histories so that we can track in detail the time-evolution of the \textit{X}-ray history beyond that driven by linear growth.

Moreover, while both codes use the global ionization fraction to evolve the attenuation of the emission spectrum, \cmfast's simplified treatment depends on a top-hat attenuation model in which previously emitted photons are either fully absorbed or unabsorbed. Our attenuation through our \dmcm transfer functions makes no such assumptions and can track the full energy-dependence of the emission spectrum as it experiences attenuation. Additionally, our transfer functions represent a wholesale improvement upon those used in \cmfast as they account for the gas density dependence in X-ray absorption.

\subsection{Global and Power Spectrum Signals}
\label{sec:PowerSpectra}

With our simulation procedure fully defined,  we proceed to consider two illustrative examples. We consider the decay of relatively light DM with $m_\chi = 5\,\mathrm{keV}$ to photons with a lifetime of $\tau = 10^{25}\,\mathrm{s}$ and heavier DM with $m_\chi = 100\,\mathrm{MeV}$ decaying to electrons with $\tau = 10^{25}\,\mathrm{s}$. These scenarios are marginally compatible with cosmological constraints from the Ly$\alpha$ forest and the CMB power spectrum.
\begin{figure}[!t]  
    \hspace{0pt}
    \vspace{-0.2in}
    \begin{center}
    \includegraphics[width=0.49\textwidth]{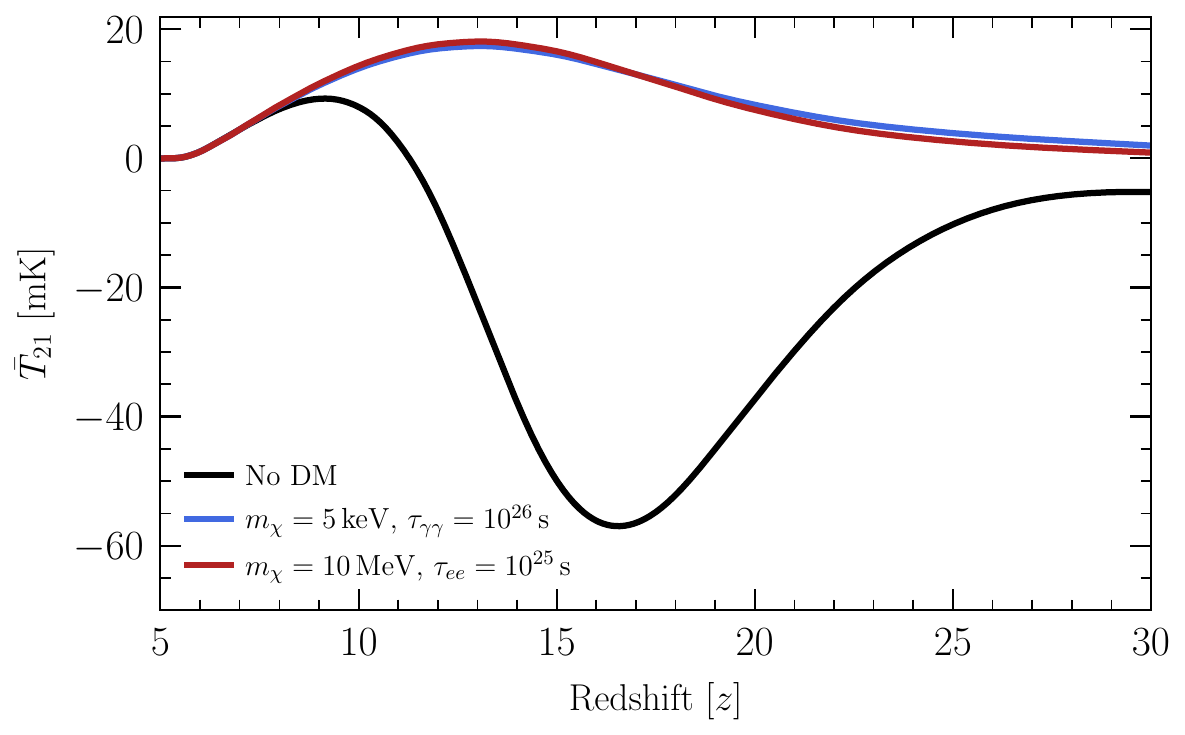}
    \caption{\textbf{Example global brightness temperature $\bar T_{21}$ evolution.} The red line shows the case of DM decay to photons with $m_\chi = 5\,\mathrm{keV}$, and the blue line shows the case of DM decay to electrons with $m_\chi = 100\,\mathrm{MeV}$. In both scenarios, the DM lifetime is taken to be $10^{25}\,\mathrm{s}$ as an illustratively large decay rate for comparison with the no DM scenario in black.}
    \label{fig:GlobalEvolution}
    \end{center}
\end{figure}

First, in Fig.~\ref{fig:GlobalEvolution}, we demonstrate the evolution of the global brightness temperature $\bar T_\mathrm{21}$ for the two scenarios, contrasting these results with $\bar T_\mathrm{21}$ due solely to the background astrophysical processes modeled by \cmfast, i.e.\ without DM energy injection (see Sec.~\ref{sec:BkgModel} for more details). Due to appreciable heating of the baryons via DM decay, the kinetic temperature lies above the CMB temperature throughout the dark ages for these DM scenarios, leading to a positive brightness temperature for all times relevant for 21-cm observations ($z \lesssim 30)$. Models which are currently consistent with cosmological observables can therefore produce drastic changes to the 21-cm signal, clearly demonstrating the unprecedented sensitivity of 21-cm cosmology to DM decay. This confirms earlier studies on the sensitivity of the global signal to decays, such as in Refs.~\cite{2013MNRAS.429.1705V,Evoli_2014, Clark:2018ghm, Mitridate:2018iag, DAmico:2018sxd, Hektor:2018qqw, Liu:2018uzy}, but with a much more sophisticated analysis including astrophysical effects and inhomogeneous energy injection (both astrophysical and exotic).

\begin{figure*}[!t]  
    \hspace{0pt}
    \vspace{-0.2in}
    \begin{center}
    \includegraphics[width=0.99\textwidth]{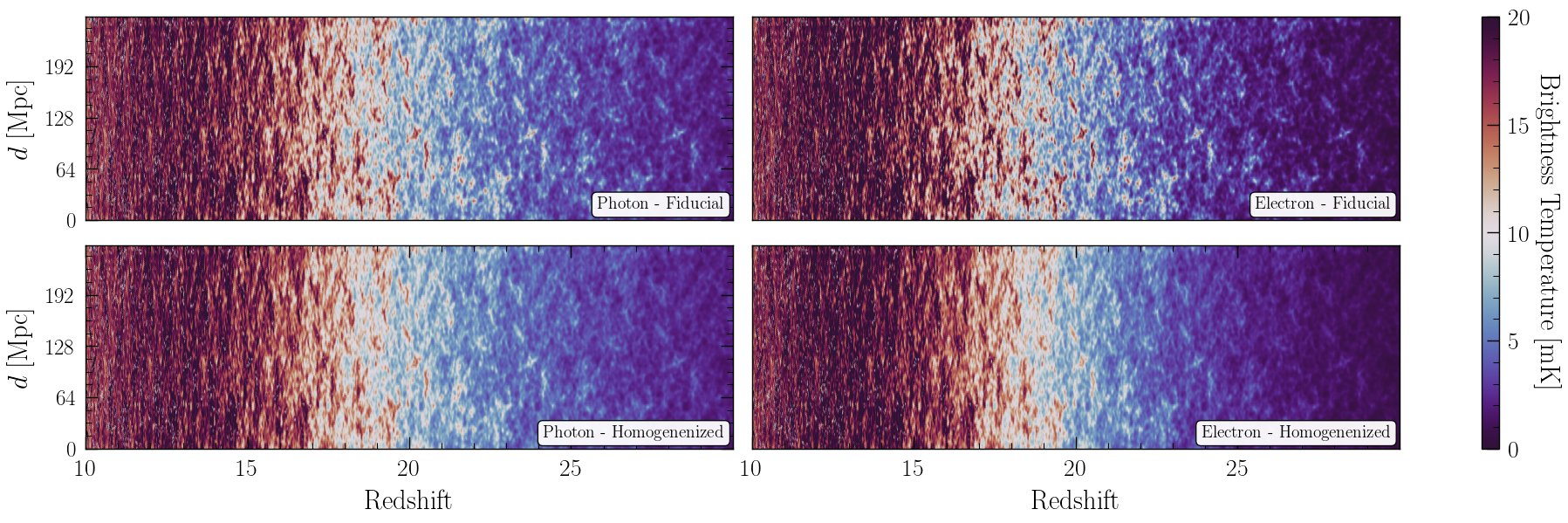}
    \caption{\textbf{Example $T_{21}$ lightcones under DM decay.} (\textit{Top Panels}) The lightcones of the brightness temperature $T_{21}$ calculated using our fiducial simulation procedure for DM decaying to photons with $m_\chi=5$~keV and a lifetime of $\tau = 10^{26}\,\mathrm{s}$ (left) and for DM decaying to electrons with $m_\chi=10$~MeV and $\tau = 10^{25}\,\mathrm{s}$ (right). (\textit{Lower Panels}) As in the upper panels, but calculated using the homogenized energy injection and deposition procedure described in Sec.~\ref{sec:PowerSpectra}. Comparing the upper and lower panels reveals the relative importance of the spatially inhomogeneous energy deposition in capturing the spatial morphology of the brightness temperature at times $z \gtrsim 15$.}
    \label{fig:Lightcones}
    \end{center}
\end{figure*}

In Fig.~\ref{fig:Lightcones}, we illustrate lightcones of the brightness temperature for the two decay scenarios, to illustrate the difference in fluctuations in $T_{21}$. For the purposes of comparison with our fiducial procedure detailed in Sec.~\ref{sec:Implementation}, we also compute lightcones using a modified version of our simulation framework in which exotic energy injection and deposition are assumed to be completely homogeneous, i.e.\ taking all injection and deposition rates calculated with \dmcm to be equal at every point in the simulation, calculated assuming $\delta \rightarrow 0$ and $x_e \rightarrow \langle x_e \rangle$. This simplified homogenized procedure approximates the treatment of Ref.~\cite{Facchinetti:2023slb}.
As expected, we observe larger fluctuations in the brightness temperature on small scales in our fiducial inhomogeneous treatment compared to the homogenized one. The spatial inhomogeneities are generally larger in amplitude for the case of decay to electrons, which is expected due to the short energy-loss path length for electrons of all energies.

\begin{figure*}[!t]  
    \hspace{0pt}
    \vspace{-0.2in}
    \begin{center}
    \includegraphics[width=0.99\textwidth]{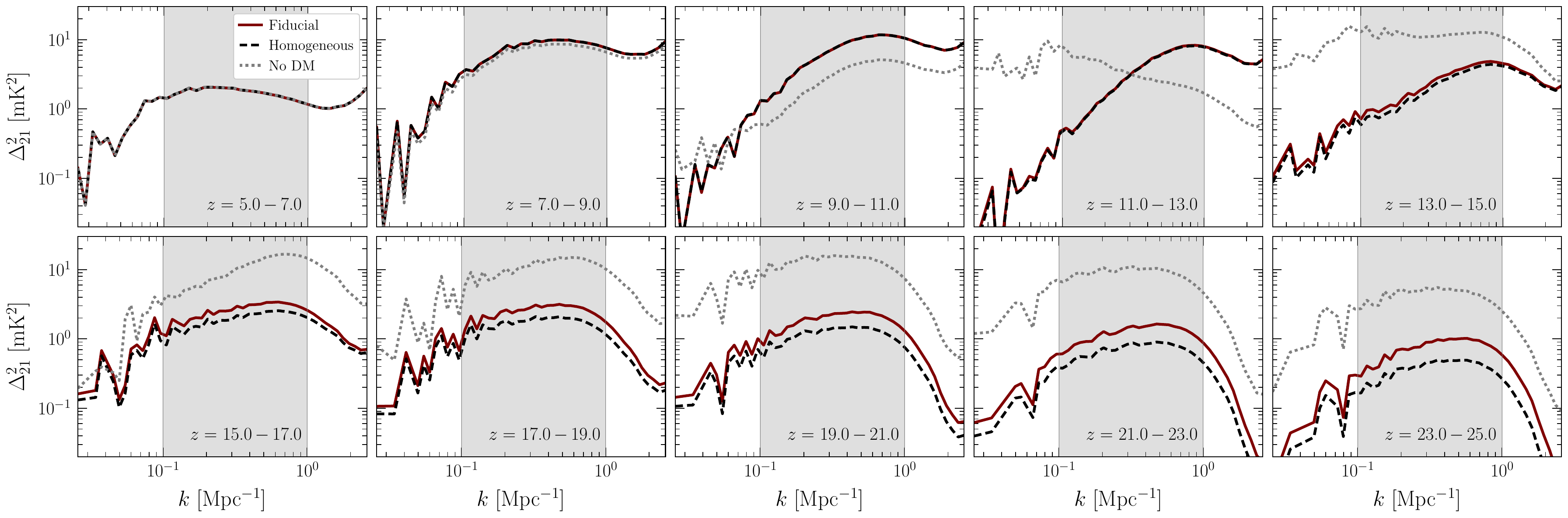}
    \caption{\textbf{Example $T_{21}$ lightcone power spectra under DM decaying to photons.} The lightcone power spectra computed for redshifts between $z=5$ and $z=25$ for the scenario of DM decay to photons for $m_\chi = 5\,\mathrm{keV}$ and $\tau = 10^{25}\,\mathrm{s}$. In maroon, we depict the power spectra calculated with our fiducial treatment while in dashed black we provide the power spectra computed with the homogenized treatment; these scenarios correspond to the upper left and lower left panels of Fig.~\ref{fig:Lightcones}, respectively. In dashed gray, we show the power spectra in the absence of DM decay. As expected, the fully inhomogeneous treatment of \textit{X}-ray emission restores some power across all scales relative to the fully homogenized treatment at early times. The grey band highlights the range of wavenumbers used for our Fisher information treatment in Sec.~\ref{sec:ProjectedConstraints}.}
    \label{fig:PhotonPowerSpectrumExamples}
    \end{center}
\end{figure*}

\begin{figure*}[!t]  
    \hspace{0pt}
    \vspace{-0.2in}
    \begin{center}
    \includegraphics[width=0.99\textwidth]{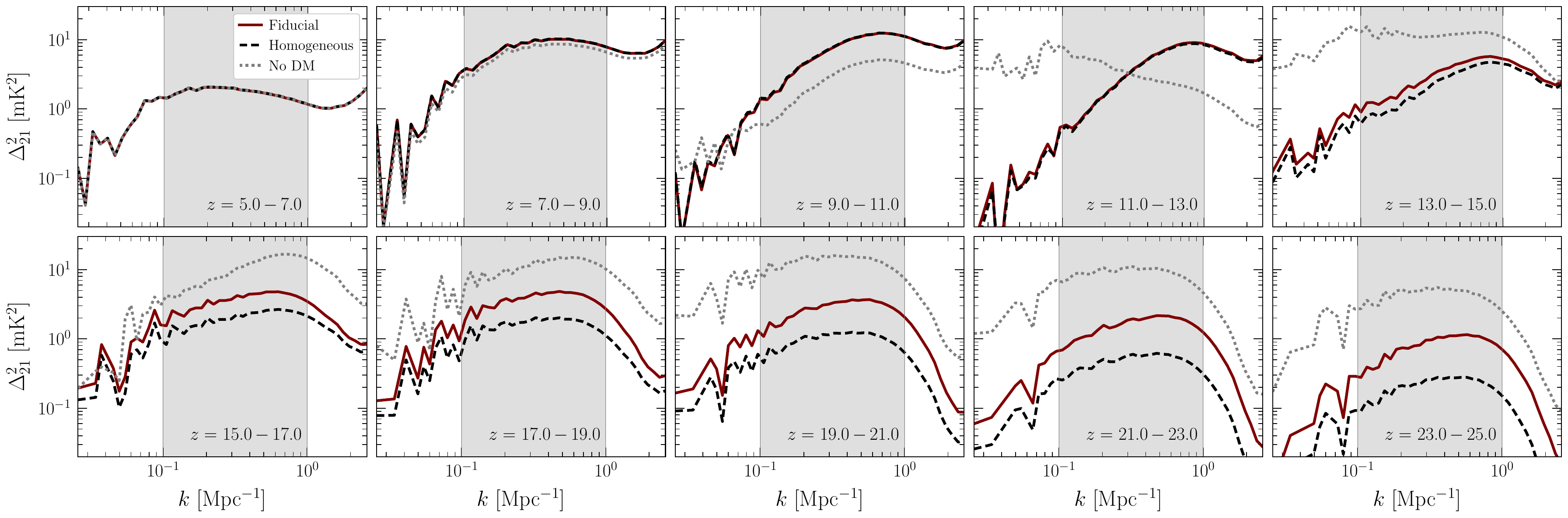}
    \caption{\textbf{Example $T_{21}$ lightcone power spectra under DM decaying to electrons.} As in Fig.~\ref{fig:PhotonPowerSpectrumExamples}, but for DM of mass $m_\chi = 10\,\mathrm{MeV}$ decaying to electrons with a lifetime of$\tau = 10^{25}\,\mathrm{s}$ as illustrated in top right and bottom right panels of Fig.~\ref{fig:Lightcones}. The considerably more ``on-the-spot" nature of energy deposition via high energy electrons leads to a more drastic difference between the fiducial and homogenized scenarios as compared to the difference observed for decay to photons.}
    \label{fig:ElectronPowerSpectrumExamples}
    \end{center}
\end{figure*}

Using these lightcones, we calculate the 21-cm power spectrum at redshifts between $z = 5$ and $z = 25$, depicted in Fig.~\ref{fig:PhotonPowerSpectrumExamples} and Fig.~\ref{fig:ElectronPowerSpectrumExamples}. These power spectra validate the observation from Fig.~\ref{fig:Lightcones} that the inhomogeneities that arise are generally larger in the decay to electron scenario relative to homogenized treatment as compared to in the decay to photons scenario. In App.~\ref{app:ExtendedLightCones}, we expand on these results, providing lightcones calculated in the absence of background astrophysics, allowing for a clearer identification of the effects of DM decay. We also independently homogenize emission and deposition to reveal the relative significance of each.

\subsection{Computational Footprint}

We also comment briefly on the computational performance of \dmcm. \dmcm uses \texttt{JAX} \cite{jax2018github}, which supports just-in-time compilation and vectorization of operations that takes advantage of parallelization on CPUs and GPUs, which \dmcm has made liberal use of. Notably, the computationally expensive operations implemented in \dmcm (Fourier transformation and linear interpolation over large look-up tables) are considerably accelerated when run on GPUs with up to a factor of 100 speedup, which uniquely enable this study. Indeed, all calculations in this paper were run using single compute nodes with 32 CPUs and 1 A100 GPU, with the \cmfast calls taking $\sim10$~s per step (over 100 steps), and the remainder of the \dmcm routine taking $\sim1$~s per subcycle step (over 1000 subcycle steps) at our fiducial simulation volume and resolution. The efficiency of \cmfast does not appear to scale as expected with increased parallelization~\cite{Murray:2020trn}; we anticipate a \texttt{JAX}-based and GPU-accelerated implementation could realize significant improvements in performance that would broadly enable more detailed modeling and higher resolution analyses beyond the DM context considered here.

\section{21-cm Sensitivity to Monochromatic Decays}
\label{sec:Proj}

In this section, we perform analyses with mock datasets and make projections for HERA sensitivities to two decaying dark matter scenarios: monochromatic decays to photons and monochromatic decays to an $e^+e^-$ pair. In both channels, we incorporate the full stellar energy injection model as implemented in \cmfast alongside our treatment of exotic energy injection from DM. The stellar \textit{X}-ray and UV radiation, especially occurring at early times ($z \gtrsim 15$), represents a confounding background with its parametrization representing nuisance parameters that weaken the expected sensitivity to, \textit{e.g.}, the DM decay rate. In Sec.~\ref{sec:BkgModel}, we give a brief overview of this background model before developing projected constraints across a range of masses, developed using our full simulation framework as an input for the \texttt{21cmfish} forecasting tool~\cite{Mason:2022obt}, in Sec.~\ref{sec:ProjectedConstraints}.

While we have chosen to consider only monochromatic decays, alternative scenarios can be straightforwardly incorporated via modification of the spectrum of injected photons and electrons. Similarly, annihilation with its less trivial density dependence may be accommodated by modifying the dependence of photon and electron emission on the local density contrast $\delta$. While annihilation---both velocity dependent and independent---is also an interesting scenario to study, the energy injection rate and spectrum may be dominated by annihilation in halos, and therefore require additional subgrid physics modeling. We leave a detailed study of the annihilation signal to future work.

\subsection{Overview of Background Modeling Parametrization}
\label{sec:BkgModel}
We provide a more complete description of the current modeling of standard astrophysical processes by \cmfast in App.~\ref{app:PressSchech}, while in this section, we summarize the salient details of the parameterization of Ref.~\cite{Qin:2020xyh} as it informs our Fisher forecasts. Just as in the DM scenario we have considered thus far, stellar emission of UV and \textit{X}-ray photons leads to energy deposition into heating, ionization, and Ly$\alpha$ excitation. These effects then drive the time evolution of the kinetic temperature $T_k$, the ionization fraction $x_e$, and the spin temperature $T_S$. \cmfast models the UV and \textit{X}-ray emission as proportional to the SFRD, meaning that these effects become important when the first halos that are large enough to host stars form. Using the built-in parametrization of \cmfast, we consider two stellar populations, which we refer to as population II and population III (PopII and III, respectively).  

We assume that PopIII stars reside within the first-forming molecular cooling galaxies (MCGs), while PopII stars are found within the later-forming atomic cooling galaxies (ACGs). MCGs and ACGs appear at different times in the cosmological history and vary in terms of size, virial temperature, and atomic/molecular composition. As a result, the two stellar populations have distinct star-formation histories and are expected to differ in terms of their luminosities relevant to the reionization process.

PopII and PopIII star-formation efficiencies are described by the population-specific parameters $\{f_{\star,10}^\text{II}, \alpha_{\star}^\text{II}, f_\mathrm{esc, 10}^\text{II}, L_X^\text{II}\}$ and $\{f_{\star,7}^\text{III}, \alpha_{\star}^\text{III}, f_\mathrm{esc, 7}^\text{III}, L_X^\text{III},  A_\mathrm{LW}\}$, respectively, and the shared parameters $\{t_{\star}, \alpha_\mathrm{esc}, E_0\}$. The Lyman-Werner feedback on MCGs, affecting the PopIII star formation, is additionally described by the parameter $A_\mathrm{LW}$~\cite{Machacek:2000us}. For a given population, $f_\star$ and $\alpha_\star$ determine the stellar mass fraction as a function of halo mass while $t_\star$ sets the star formation rate as a function of the stellar mass and Hubble time. The parameters $f_\mathrm{esc}$ and $\alpha_\mathrm{esc}$ determine the escape fraction of UV radiation, which sets the efficiency of galaxies in reionizing hydrogen within their vicinity, while $L_X$ and $E_0$ set the \textit{X}-ray luminosity relative to the SFRD and the minimum energy of propagating \textit{X}-rays \cite{Park:2018ljd, Munoz:2021psm}. 

\begin{table}[!ht]{
    \ra{1.3}
    \begin{center}
    \tabcolsep=0.08cm
    \begin{tabular}{c*{4}{C{0.07\textwidth}}}
    \hlinewd{1pt} 
    \textbf{PopII} parameters & $f_{\star,10}^\text{II}$ & $\alpha_\star^\text{II}$ & $f_\mathrm{esc,10}^\text{II}$ &  $L_X^\text{II}$ \\
    Fiducial value & -1.25 & 0.5 & -1.35 & 40.5 \\ \hlinewd{0.5pt}
    \textbf{PopIII} parameters & $f_{\star,7}^\text{III}$ & $\alpha_\star^\text{III}$ & $f_\mathrm{esc,7}^\text{III}$ &  $L_X^\text{III}$ \\ 
    Fiducial value & -2.5 & 0.0 &-1.35 & 40.5 \\ \hlinewd{0.5pt}
    \textbf{Shared} parameters & $t_\star$ & $\alpha_\mathrm{esc}$ & $E_0$ &  $A_\mathrm{LW}$ \\
    Fiducial value & 0.5 & -0.3 & 500 & 2.0 \\ \hlinewd{1.0pt}
    \end{tabular}\end{center}}
\caption{\textbf{Summary of nuisance parameters in Fisher forecast.} The nuisance parameters describing the star formation rate density and associated \textit{X}-ray luminosity in \cmfast with fiducial values adopted from Ref.~\cite{Munoz:2021psm, Mason:2022obt}. In our Fisher information treatment using \texttt{21cmfish}, each parameter is independently varied. For details, see Sec.~\ref{sec:BkgModel} and Refs.~\cite{Qin:2020xyh, Munoz:2021psm}. 
\label{tab:FisherNuisanceParameters}}
\end{table}

We adopt the fiducial parametrization associated with the ``best-guess" scenario for 21-cm power spectrum modeling with \cmfast developed in Ref.~\cite{Munoz:2021psm} and studied in a Fisher forecast using \texttt{21cmfish} in Ref.~\cite{Mason:2022obt}. Using the \texttt{21cmfish} framework, we develop projected sensitivities for the 21-cm power spectrum at comoving wavenumbers between $0.1 \, \mathrm{Mpc}^{-1}$ and $1.0 \, \mathrm{Mpc}^{-1}$ as is expected to be measured by HERA, using 331 antennae for a total exposure of 1080 hours at $8\,\mathrm{MHz}$ radio frequency bandwidths between $50$ and $250\,\mathrm{MHz}$. Like in the treatment of \texttt{21cmfish}, we make use of the moderate foreground model developed in \texttt{21cmSense} and assume a HERA system temperature 
\begin{equation}
T_\mathrm{sys} = 100 \,\mathrm{K} + 120 \,\mathrm{K} \times \left(\frac{\nu}{150 \,\mathrm{MHz}} \right)^{-2.55},
\end{equation}
where $\nu$ is the observation frequency. We also include Poisson uncertainty and a 20\% modeling systematic uncertainty in our error budget following Ref.~\cite{Park:2018ljd}. These choices represent the astrophysics and uncertainty modeling used in the Fisher forecast of Ref.~\cite{Mason:2022obt} to reproduce the Bayesian analysis of Ref.~\cite{Munoz:2021psm}.

\subsection{Projected Constraints on Dark Matter Decays}
\label{sec:ProjectedConstraints}

\begin{figure*}[t]
    \hspace{0pt}
    \vspace{-0.2in}
    \begin{center}
    \includegraphics[width=0.99\textwidth]{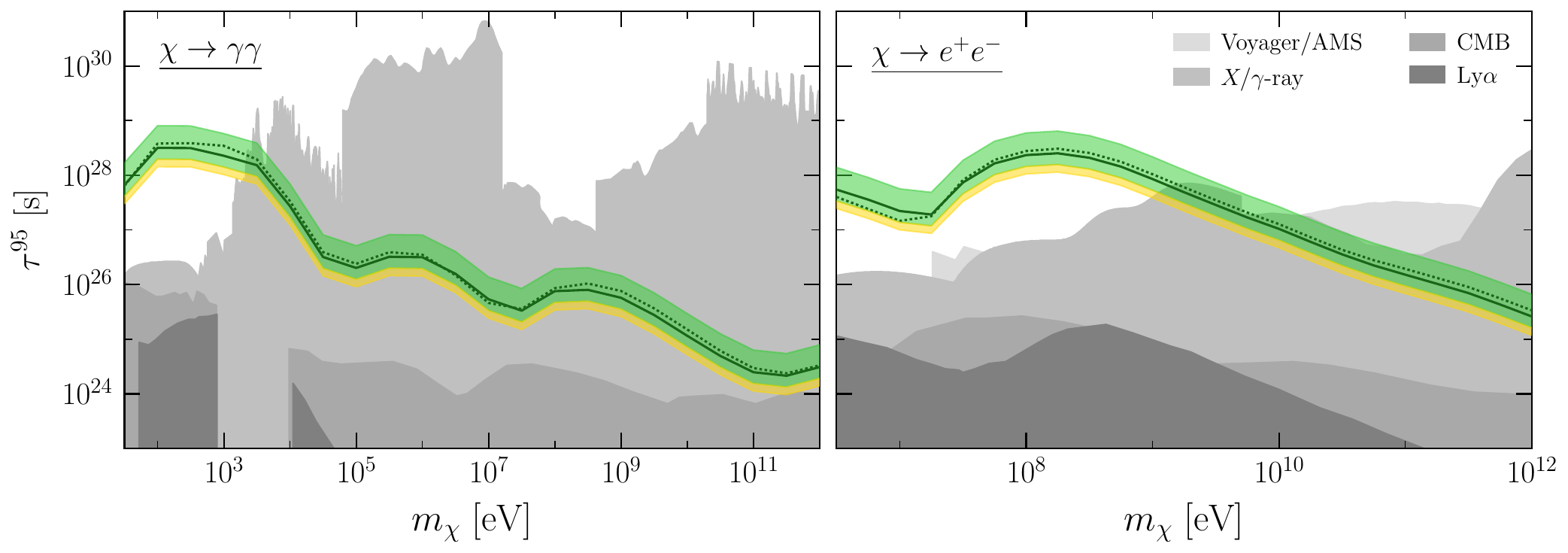}
    \caption{\textbf{Projected 95$^\text{th}$ percentile lower limits on monochromatic decays.} (\textit{Left}) Limits on monochromatic decays to photons (solid black) with the 1(2)$\sigma$ containment intervals of the expected limit indicated by the green (yellow) band. Shaded grey regions depict existing bounds on DM decay to photons provided from the CMB~\cite{Slatyer:2016qyl, capozzi2023cmb}, the Ly$\alpha$ forest~\cite{liu2021lyman, capozzi2023cmb}, heating of gas-rich dwarf galaxies~\cite{Wadekar:2021qae}, and \textit{X}-ray and $\gamma$-ray observations~\cite{Arias:2012az, Essig:2013goa, Horiuchi:2013noa, Ng:2019gch, Sicilian:2020glg, Foster:2021ngm, Foster:2022nva, Roach:2022lgo,Cirelli:2023tnx}. A dotted black line indicates the projected limits calculated using the homogenized energy emission and deposition treatment. (\textit{Right}) As in the left panel, but for monochromatic decays to $e^+e^{-}$. An additional source of constraints on this parameter space is provided by Voyager and AMS measurements of charged cosmic rays~\cite{Massari:2015xea, Giesen:2015ufa,Boudaud:2016mos}.}
    \label{fig:ProjectedLimits}
    \end{center}
\end{figure*}

To develop projected constraints across a broad range of masses for the scenarios of DM decays to photons and to electrons, we generate an ensemble of 21-cm power spectra calculated with and without exotic energy injection. For decay to photons, we consider a range of masses $m_\chi$ from $100\,\mathrm{eV}$ to $1\,\mathrm{TeV}$; we consider masses between $10\,\mathrm{MeV}$ and $1\,\mathrm{TeV}$ for decay to electrons. Our simulations are performed using a box of comoving volume $(256 \,\mathrm{Mpc})^3$ resolved by $128^3$ lattice sites. Treating each simulated mass independently, we use $\texttt{21cmfish}$ to determine the Fisher information for the one signal parameter (the dark matter decay rate) and the twelve nuisance astrophysical background parameters. The Fisher matrix is calculated by first performing simulation under a fiducial parametrization that provides a mock dataset; next, for each model parameter, two independent simulations are performed in which the model parameter of interest is varied about its fiducial value. This allows us to compute the derivatives of the likelihood around our fiducial (which is assumed to maximize the likelihood by construction), and thus to compute the Fisher information matrix. Assuming parameter sensitivity at the Cramer-Rao bound, we use the Fisher matrix to determine the projected frequentist 95$^\mathrm{th}$ percentile upper limits on the DM decay rate (equivalently, 95$^\mathrm{th}$ percentile lower limits on the DM lifetime) \cite{Cowan:2010js}.

The projected limits on DM decay to photons and electrons are shown in the left and right panels of Fig.~\ref{fig:ProjectedLimits}, respectively. The projected constraints from the decay to photons surpass at all masses the Ly$\alpha$ and CMB constraints and exceed \textit{X}-ray line constraints at masses below $1\,\mathrm{keV}$. In the case of decay to electrons, the projected constraints from the 21-cm power spectrum are substantially stronger than those realized by Ly$\alpha$ and CMB constraints and would represent the strongest constraint on particles decaying to electrons for masses below $10\,\mathrm{GeV}$.

The simulation procedure we have developed in this work is fundamentally motivated by the expectation that the fundamentally inhomogeneous process of DM-sourced energy injection and deposition would leave a distinct imprint upon the 21-cm power spectrum, and we have indeed found this to be true, as shown in Fig.~\ref{fig:lightconediff} and studied in Sec.~\ref{sec:PowerSpectra}. It is then interesting to examine how these projected sensitivities differ to those computed without a proper treatment of inhomogeneities in DM processes, such as performed in Ref.~\cite{Facchinetti:2023slb}. We note that our projected limits here cannot be directly compared to those of Ref.~\cite{Facchinetti:2023slb}, which makes use of a different astrophysics and noise modeling than developed in Refs.~\cite{Munoz:2021psm, Mason:2022obt}. To then enable a more direct assessment of the importance of modeling the inhomogeneities, we also develop projected limits with the signal calculated using our simplified homogenized treatment described in Sec.~\ref{sec:PowerSpectra}.

Surprisingly, we find that the projected sensitivities calculated with the homogenized treatment are not appreciably different and in some cases are stronger at the $\mathcal{O}(10\%)$ level than those calculated with the fully inhomogeneous treatment. In fact, this could have been anticipated by observing that both DM and stellar reionization processes track the density contrast field. As a result, stronger limits are generated by the homogenized treatment, which predicts a DM signal with less degeneracy with the standard astrophysics signal. Given this, it is encouraging that the limits are only marginally weakened when calculated by a correctly inhomogeneous modeling procedure. 

We caution against interpreting similar projected limits as evidence that the inhomogeneities in DM processes are unimportant. As is immediately clear in Fig.~\ref{fig:lightconediff}, the two methods of calculation make predictions that for $T_{21}$ that differ appreciably in their small-scale power, making the inhomogeneous and homogenized scenarios distinguishable from one another, if also comparably distinguishable from the null hypothesis of no DM decay. If the 21-cm power spectrum does indeed contain evidence for exotic energy injection, the more accurate modeling framework we have developed here will be critical for the accurate modeling and interpretation of a potential discovery. Moreover, \texttt{21cmfish} makes considerable simplifications in its noise modeling by treating the measurement uncertainty at each frequency and in each wavenumber as uncorrelated. Correlated uncertainties, which are likely to arise in real datasets, could further complicate the extraction of a DM signal, making the robust and realistic framework developed here especially important.

\subsection{Triangle Plots}

\begin{figure*}[!t]  
    \hspace{0pt}
    \vspace{-0.2in}
    \begin{center}
    \includegraphics[width=0.99\textwidth]{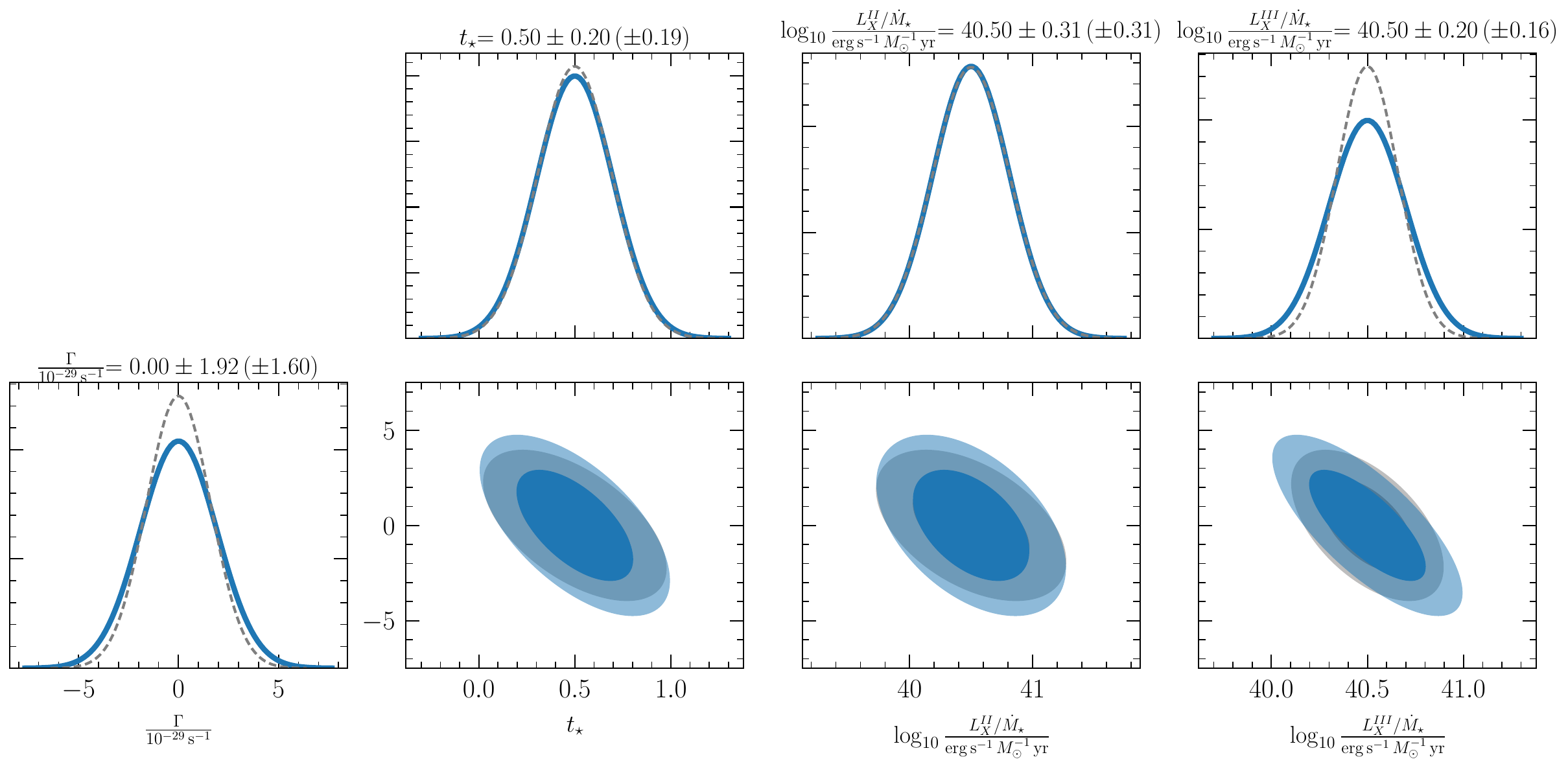}
    \caption{\textbf{Covariance between decay rate and the most correlated  astrophysical parameters in $\chi\rightarrow\gamma\gamma$.} The subset of the triangle plots demonstrating the parameter sensitivities to and covariances between the DM decay rate $\Gamma$ and the astrophysical parameters with which it is most correlated: the star formation rate parameter $t_\star$ and the PopII and PopIII luminosity parameters $L_X^\text{II}$ and $L_X^\text{III}$. In this particular scenario, we have taken $m_\chi = 100\,\mathrm{eV}$ where we achieve the strongest projected constraints on decay to photons. The 1(2)$\sigma$ confidence intervals are shown in dark (light) blue shading for the fiducial, inhomogeneous treatment while corresponding contours are shown in gray for the homogenized treatment. The individual $1\sigma$ parameter uncertainties are also provided for the fiducial treatment and parenthesized for the homogenized treatment in blue and dashed grey, respectively.} 
    \label{fig:PhotonTriangleSubset}
    \end{center}
\end{figure*}

\begin{figure*}[!t]  
    \hspace{0pt}
    \vspace{-0.2in}
    \begin{center}
    \includegraphics[width=0.99\textwidth]{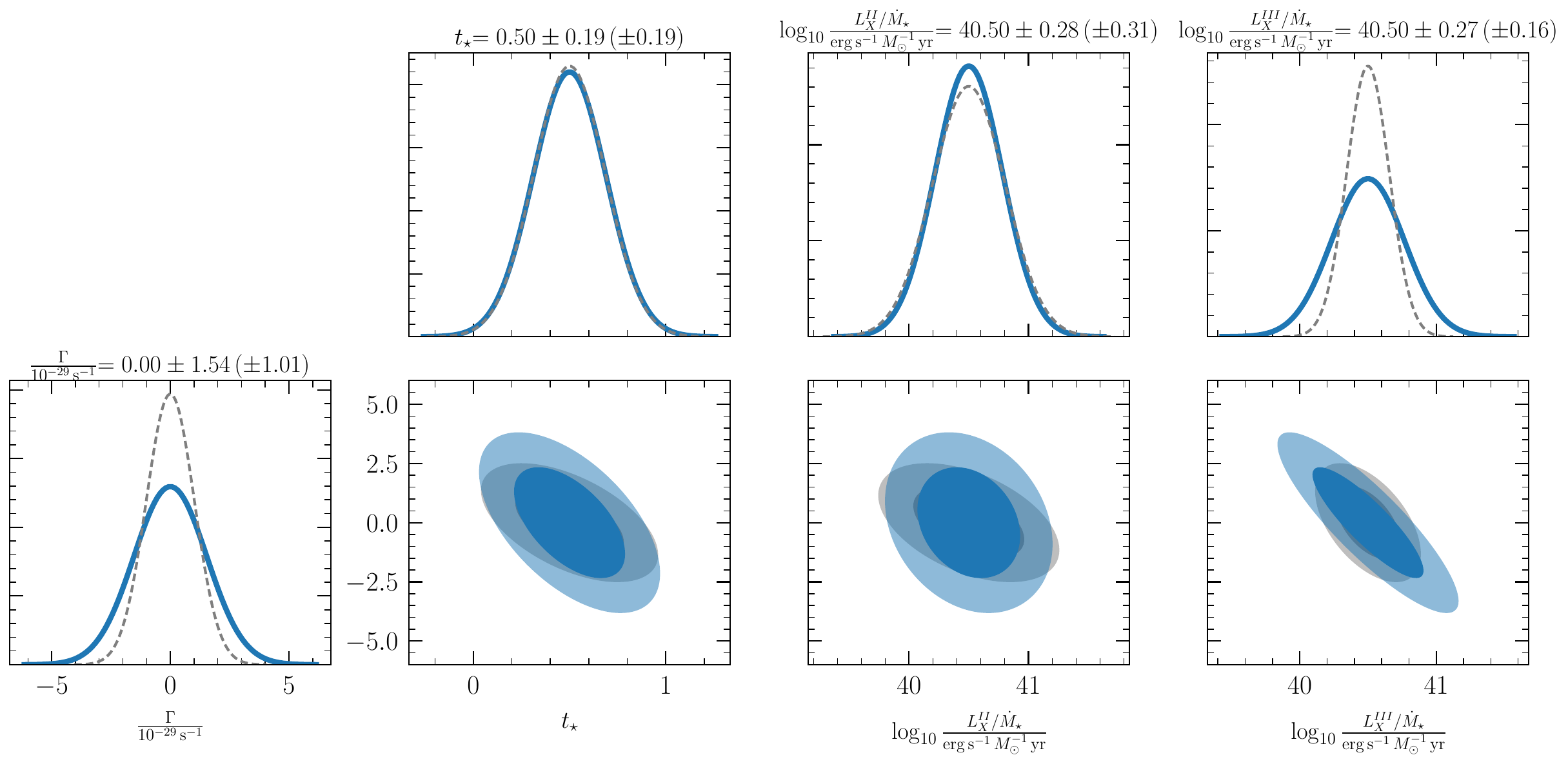}
    \caption{\textbf{Covariance between decay rate and the most correlated astrophysical parameters in $\chi\rightarrow e^-e^+$.} As in Fig.~\ref{fig:PhotonTriangleSubset}, but for the scenario of electron decay. As before, we choose the DM mass associated with our strongest projected constraint, in this case $m_\chi = 100\,\mathrm{MeV}$.}
    \label{fig:ElectronTriangleSubset}
    \end{center}
\end{figure*}

Given the considerable astrophysical uncertainties associated with modeling the 21-cm power spectrum, it is informative to inspect the estimated parameter covariances obtained through our Fisher information treatment. In general, we find estimates of the DM lifetime $\tau$ are mostly correlated with estimates of the star formation parameter $t_\star$ and the PopII and PopIII luminosity parameters $L_X^\text{II}$ and $L_X^\text{III}$. We present the $1\sigma$ and $2\sigma$ confidence intervals on the one-dimensional space for each of these parameters and the two-dimenisonal space of their pairwise combinations in the triangle plots for decay to photons in Fig.~\ref{fig:PhotonTriangleSubset} and decay to electrons in Fig.~\ref{fig:ElectronTriangleSubset}. These contours are illustrated for the masses which achieve the strongest constraints for a given decay channel: $m_\chi = 100\,\mathrm{eV}$ for decay to photons and $m_\chi = 100\,\mathrm{MeV}$ for decay to electrons. We also compare the confidence intervals obtained under our fiducial inhomogeneous treatment with those obtained with our homogenized treatment. 

In the case of decays to photons shown in Fig.~\ref{fig:PhotonTriangleSubset}, and decay to electrons in Fig.~\ref{fig:ElectronTriangleSubset}, we find that the correlation between the DM decay rate $\Gamma$ and the star formation rate parameter $t_\star$ and PopII luminosity parameter $L_X^\text{II}$ are not appreciably different in the fiducial inhomogeneous calculation as compared to the simplified homogenized calculation. On the other hand, accounting for inhomogeneities increases the uncertainty in the PopIII luminosity parameter $L_{X}^\text{III}$. This owes to the inhomogeneous DM processes being more similar to the \textit{X}-ray emission from early star formation, and thus more degenerate.  The full triangle plots for all 13 parameters considered in our analysis subject to our choice of representative mass are provided in App.~\ref{app:FullTriangle}.

\section{Conclusion}
\label{sec:Conclusion}
In this paper, we present \dmcm, a self-consistent computation of the 21-cm signal in the presence of exotic energy injection, properly accounting for both inhomogeneous injection and deposition. Our code is publicly available, and our calculation framework is compatible with existing tools for 21-cm power spectrum analyses and projections, such as \texttt{21cmfish} and \texttt{21CMMC}. We have used this new \dmcm framework to make robust predictions for the sensitivity of the 21-cm power spectrum to decaying DM. We find strong projected limits, which both for decay to photons and electron-positron pairs can outperform current limits in different mass ranges.
Importantly, our estimated sensitivities account for the first time for the effect of inhomogeneities in the DM processes,  which is critical to modeling the 21-cm power spectrum accurately and obtaining robust limits.

Our results largely validate the use of a homogenizing approximation in the previous literature \cite{Facchinetti:2023slb}, where the energy deposition is calculated using \dhis based on the global average $x_e$ and matter density, to estimate constraints on DM decay. This approximation works reasonably well in the context of both the global signal and projected constraints from the power spectrum as measured by HERA. However, we caution that our work reveals significant differences in the effect on the redshift-dependent power spectrum between the full calculation and the homogeneous approximation, which may be relevant for experiments with different redshift-dependent sensitivity than we have assumed in forecasting HERA constraints, or when interpreting any putative signal of energy injection detected in the power spectrum.

More broadly, this work presents the first systematic study and improvement upon the treatment of energy injection originally made in the first release of \cmfast \cite{2011MNRAS.411..955M}. Indeed, while the code and associated modeling procedures have undergone substantial revisions since original publication, the energy deposition procedure has remained essentially unmodified until now. While we have used our new implementation only to incorporate the effects of energy injection from DM, our framework is a generally flexible and can accommodate a number of modified physics and cosmological scenarios. We anticipate this represents a first effort towards a more flexible, powerful, and efficient modeling procedure for 21-cm cosmology that will inform the current and coming generation of experiments.

\textbf{Note Added:} In the final stages of preparation of this manuscript, Ref.~\cite{DelaTorreLuque:2023cef} appeared, claiming strong limits on DM decays to electron/positron pairs. While systematics in the data reduction and modeling of the cosmic ray propagation have yet to be fully mapped out, those limits may attain comparable sensitivity to the ones we project here.

\begin{acknowledgments}
We thank Jordan Flitter, Laura Lopez-Honorez, Katherine Mack,  Wenzer Qin, and Tomer Volansky for helpful discussions. JF was supported by a Pappalardo fellowship. YS and TRS were supported by the U.S. Department of Energy, Office of Science, Office of High Energy Physics
of U.S. Department of Energy under grant Contract Number DE-SC0012567 through the Center for Theoretical Physics at MIT, the National Science Foundation under Cooperative Agreement PHY-2019786 (The NSF AI Institute for Artificial Intelligence and Fundamental Interactions, http://iaifi.org/), and the Simons Foundation (Grant Number 929255, T.R.S). JBM acknowledges support from the National Science Foundation under Grant No.~2307354. HL is supported by the Kavli Institute for Cosmological Physics and the University of Chicago through an endowment from the Kavli Foundation and its founder Fred Kavli, and Fermilab, operated by the Fermi Research Alliance, LLC under contract DE-AC02-07CH11359 with the U.S. Department of Energy, Office of Science, Office of High-Energy Physics. The computations in this paper were run on the Erebus machine at MIT and the FASRC Cannon cluster supported by the FAS Division of Science Research Computing Group at Harvard University. This work used NCSA Delta CPU at UIUC through allocation PHY230051 from the Advanced Cyberinfrastructure Coordination Ecosystem: Services \& Support (ACCESS) program, which is supported by National Science Foundation grants 2138259, 2138286, 2138307, 2137603, and 2138296.

This work made use of
\texttt{NumPy} \cite{harris2020array},
\texttt{Scipy} \cite{virtanen2020scipy},
\texttt{Astropy} \cite{robitaille2013astropy},
\texttt{matplotlib} \cite{hunter2007matplotlib},
\cmfast \cite{2011MNRAS.411..955M},
\texttt{21cmfish} \cite{Mason:2022obt},
\texttt{21cmSense} \cite{pober201621cmsense},
\texttt{DarkHistory} \cite{Liu:2019bbm},
\texttt{JAX} \cite{jax2018github},
and all associated dependencies.
\end{acknowledgments}
\appendix

\section{Adiabatic Evolution without Energy Injection}
\label{app:AdiabaticEvolution}
\dhis underlies \dmcm's energy injection treatment, so our first demonstration is that the global evolution of $x_e$ and $T_k$ is consistent between \dhis and \cmfast when both are used to evolve a homogeneous universe. In order to realize a homogeneous universe within \cmfast, we set the $\sigma_8$ parameter to zero, which has the additional effect of turning off any stellar UV or \textit{X}-ray emission.
We expect that in this scenario, the evolution of global quantities in \cmfast\ should match that performed by \dhis.

\begin{figure}[h!]
    \begin{center}
    \includegraphics[width=0.49\textwidth]{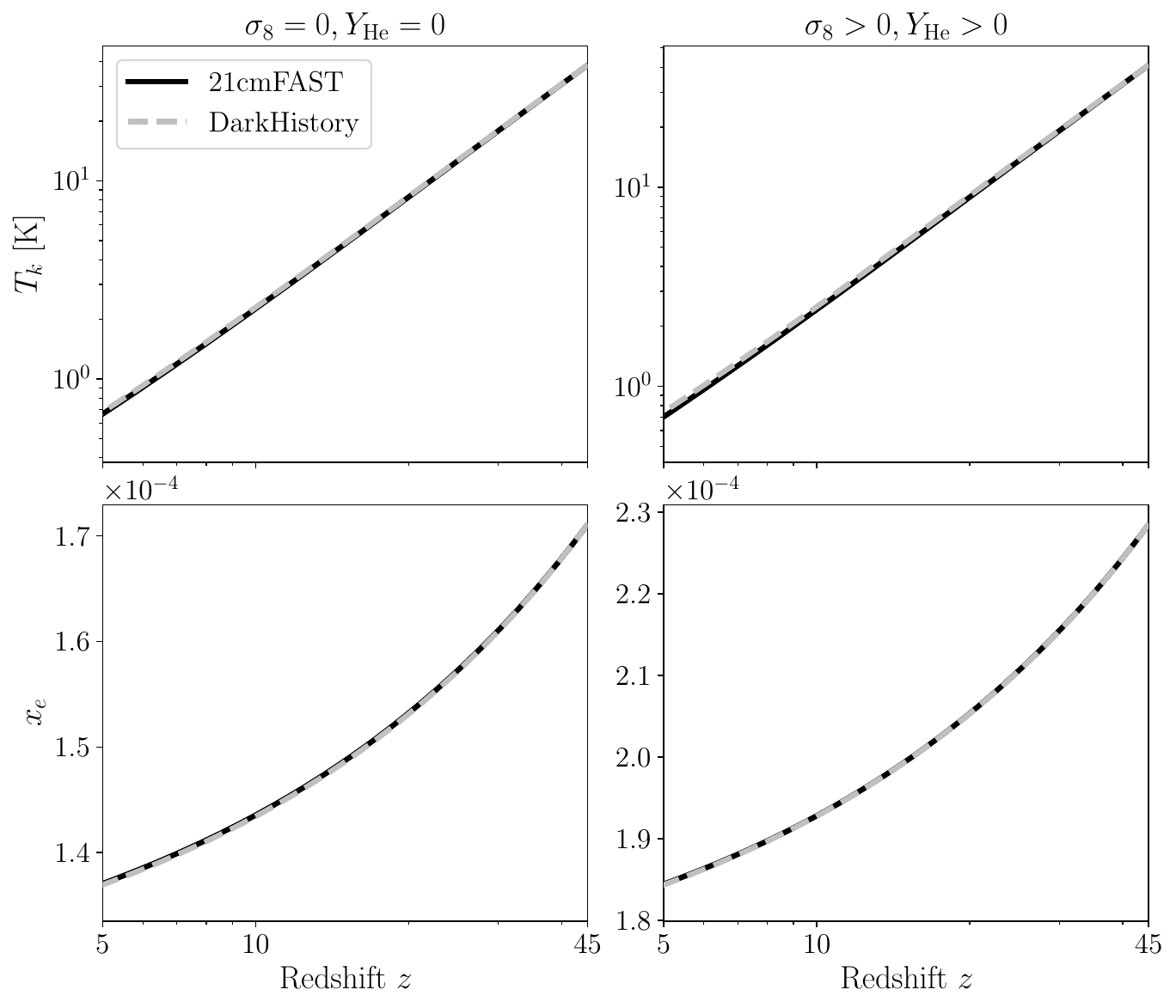}
    \caption{\textbf{Comparison of \cmfast and \dhis global evolution.} (\textit{Left column}) Kinetic temperature $T_k$ and ionization fraction $x_e$ as evolved by \dhis and \cmfast in the case of a homogeneous universe ($\sigma_8 = 0$) consisting of purely hydrogen ($Y_\mathrm{He} = 0$), with no UV or \textit{X}-ray energy injection.
    Excellent agreement is observed between the two codes, with differences attributable to finite precision in numerical integration. (\textit{Right column}) We now set $\sigma_8>0$ in \cmfast and $Y_\text{He}>0$ in both \cmfast and \dhis. This reflects the two codes in their default configuration up to the absence of \textit{X}-ray energy injection in \cmfast. Good agreement between the two codes is observed, with differences attributable to the effect of adiabatic heating and cooling due to structure formation modeled in \cmfast but not in \dhis. See text for details.}
    \label{fig:xc_21_v_DH}
    \end{center}
\end{figure}

Even at this stage, modeling differences arise. First, \dhis independently evolves the ionization of hydrogen and helium, while \cmfast takes them to be locked. For the sake of this test, we then set $Y_\mathrm{He} = 0$ in order to enforce consistency. Second, \dhis and \cmfast also implement a slightly different Compton cooling term, which we do not modify in either code. Finally, we find that the fitting formula used to calculate $dt/dz$ in \cmfast used in the forward integration realizes relative error as large as 1\%. This sets a rough floor for the best agreement we can hope to see. The time evolution of these global quantities in this case of enforced consistency is presented in the left two panels of Fig.~\ref{fig:xc_21_v_DH}.  We see excellent agreement, with maximum relative differences in the global $T_k$ and $x_e$ between \cmfast and \dhis of 2\% and 0.08\% in $T_k$ and $x_e$ respectively. 

Next, we relax the consistency conditions we have enforced. We first restore $\sigma_8$ in \cmfast to its nonzero best-fit value from the Planck 2018 analysis \cite{Planck:2018vyg} but take the \textit{X}-ray luminosity parameter $L_X$ to be zero to prevent energy injection.
We find that in this case, the maximum relative differences between the global average as evolved by \cmfast and the homogeneous universe values evolved by \dhis grow to 5\% and 0.05\% in $T_k$ and $x_e$, respectively. This difference can be attributed to the impact of adiabatic heating and cooling from structure formation, which is included in \cmfast but cannot be
incorporated in \dhis.

We continue by additionally restoring nonzero $Y_\mathrm{He}$ in both codes
while preserving vanishing $L_X$. In this comparison, in which the two codes are in their default configuration and are maximally systematically different, the maximum relative difference grows marginally to 6\% and 0.05\% in the global average of $T_k$ and $x_e$. The results of this comparison are presented in the right two panels of Fig.~\ref{fig:xc_21_v_DH}.

\section{Tests of Energy Injection from Star Formation}
\label{app:SFRD}
The most direct validation we can perform to test \dmcm against \dhis and \cmfast is to compare how each code handles the effect of energy injection through \textit{X}-ray emission tracing the SFRD. To do so, we adopt the ``mass-dependent $\zeta$'' treatment of \cmfast with default parameters to predict \textit{X}-ray emissivities.

A full accounting of the \textit{X}-ray treatment with mass-dependent $\zeta$ can be found in \cite{Park:2018ljd, Qin:2020xyh,Munoz:2021psm}, which we review here briefly. To calculate the incident \textit{X}-ray flux, for each location in the simulation volume, an extended Press-Schechter scheme is used to evaluate a halo mass function evaluated on the lightcone. This halo mass function is then integrated assuming a mass-dependent relationship for the efficiency of star formation in a halo. Under a simple parametrization, this enables a calculation of the SFRD and associated \textit{X}-ray luminosity along the lightcone. Though each pixel in the simulation has an incident flux obtained by independent lightcone integrals of the luminosity, at fixed $z$ on a lightcone, the average luminosity is normalized to the \textit{X}-ray luminosity associated with the Sheth-Tormen prediction for the halo mass function. The key points are (1) that the total \textit{X}-ray emission is fixed by construction at each $z$ along the lightcone to match the Sheth-Tormen prediction, and (2) that the extended Press-Schechter calculation along the lightcone is performed by evaluating the density contrast field at past redshifts.

\subsection{Tests of Energy Injection against \dhis}

\begin{figure}[!t]
    \begin{center}
    \includegraphics[width=0.49\textwidth]{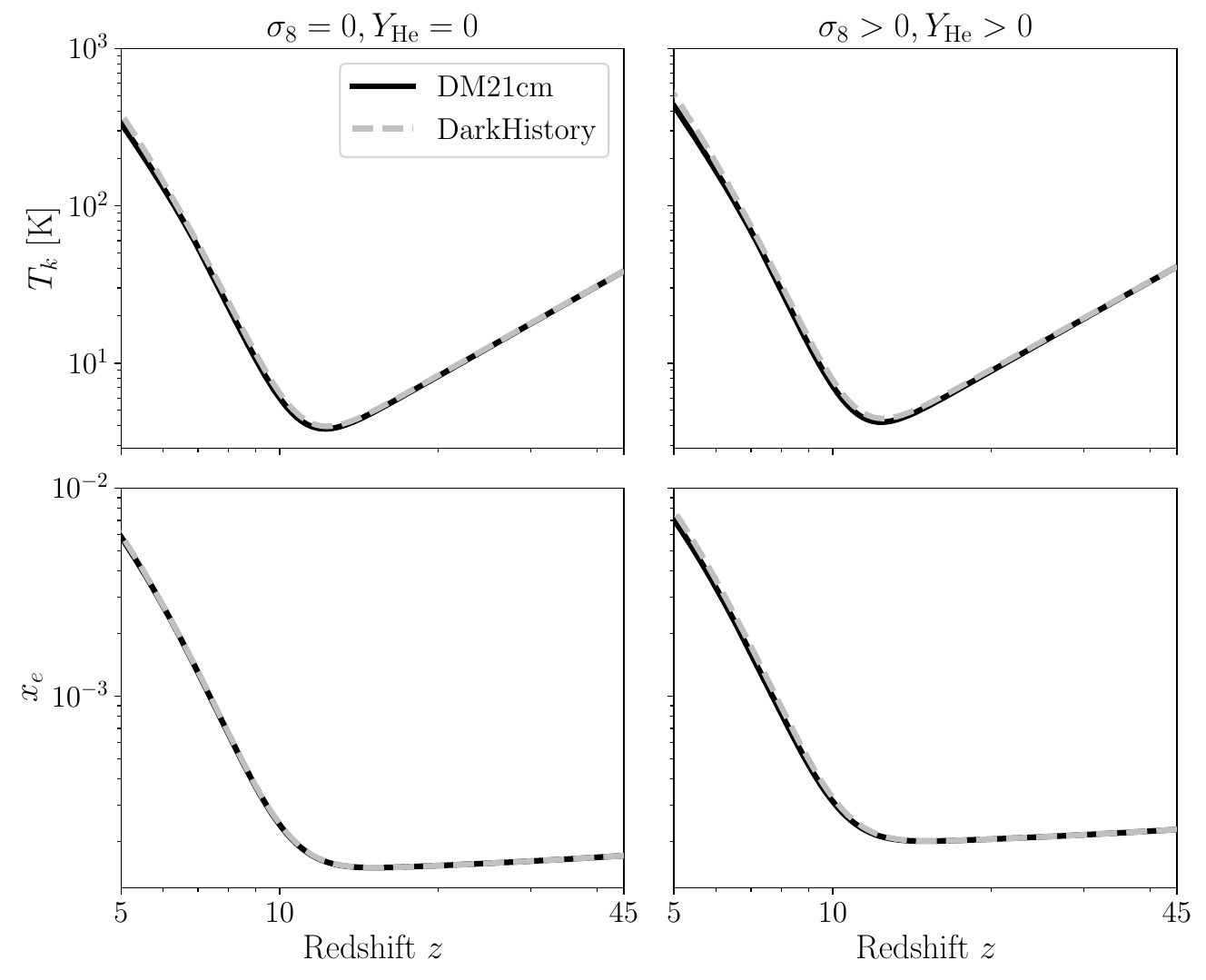}
    \caption{\textbf{Comparison of \dmcm and \dhis global evolution under star formation \xray injection.} (\textit{Left column}) \xray injection in a homogeneous universe assuming unconditional Sheth-Tormen halo mass function, with $Y_\text{He}=0$ to match \dmcm and \dhis as closely as possible. 
    (\textit{Right column}) \xray injection in a inhomogeneous universe assuming a hybrid conditional halo mass function in \dmcm, compared against homogeneous injection in \dhis. We also restored the correct value of $Y_\text{He}$. See text for details.}
    \label{fig:xc_xrayVDH_nos8_noHe_nosp}
    \end{center}
\end{figure}

In Sec.~\ref{app:AdiabaticEvolution}, we have established that \cmfast and \dhis realize nearly identical adiabatic evolution of the quantities $T_k$ and $x_e$ absent energy injection. Since \dmcm inherits its adiabatic evolution from \cmfast and its energy injection from \dhis, a straightforward consistency test is to compare the treatment of homogeneously injected \textit{X}-rays processed through the full framework of \dmcm with the treatment of injected \textit{X}-rays in \dhis.

For this test, we first take $Y_\mathrm{He} = 0$ to make the adiabatic evolution as identical as possible. We then step along in $z$, injecting \textit{X}-rays from a spatially homogeneous emission with total luminosity set by the unconditional Sheth-Tormen prediction for the globally averaged halo mass function and associated SFRD. We emphasize that although the input emission is spatially homogeneous, it is processed through the full \textit{X}-ray photon framework of \dmcm, which is unaware of this underlying homogeneity. We also perform a pure \dhis evaluation using an identical input \textit{X}-ray emission. 

We then compare the evolution of the globally averaged $T_k$ and $x_e$ obtained by \dmcm and \dhis for these two runs. The results are presented in Fig.~\ref{fig:xc_xrayVDH_nos8_noHe_nosp}. We find a maximum relative difference of 5\% in $T_k$ and 7\% in $x_e$ between the two runs. Given that \dmcm and \dhis realize nearly identical adiabatic evolution in the $\sigma_8,\, Y_\mathrm{He} \rightarrow 0$ limit and \dmcm inherits its energy deposition treatment from \dhis, the size of this discrepancy may seem somewhat surprising. We have found that the precision loss in this case is primarily driven by the finite resolution of the precomputed transfer functions obtained from \dhis used in \dmcm. 

We further extend this test by restoring $\sigma_8$ and $Y_\mathrm{He}$ in \dmcm though we now disable the \textit{X}-ray and UV luminosity, which has the effect of fully eliminating the ionizing and heating effects of stellar evolution while allowing for an inhomogeneous universe. As our universe is inhomogeneous, we now allow for our \textit{X}-ray emission to be inhomogeneous by reproducing an identical mass-dependent $\zeta$ treatment to that built into \cmfast in which the local \textit{X}-ray flux is calculated from a conditional Press-Schechter halo mass function and globally normalized to the predictions of the unconditional Sheth-Tormen halo mass function. However, our treatment differs in that rather than using a linear growth factor argument to evaluate overdensity fields at prior $z$, we use our caching system to cache and access density fields previously evaluated during the \dmcm stepping. This most closely replicates our default \textit{X}-ray treatment for exotic energy injection and so is a valuable test. As before, we compare this to the homogeneous universe evolution performed in \dhis, with results presented in the right column of Fig.~\ref{fig:xc_xrayVDH_nos8_noHe_nosp}.

The maximum relative differences obtained between an inhomogeneous universe \dmcm and a homogeneous universe \dhis are 18\% in $T_k$ and 8\% in $x_e$. While the change in precision in $x_e$ as compared to that evaluated for a homogeneous universe in \dmcm is negligible, the relative difference in $T_k$ has more than doubled. While these runs do differ as \dmcm inherits heating and cooling from structure formation effects as well as a different treatment of helium as compared to \dhis, we have found that enforcing $Y_\mathrm{He} = 0$ and disabling these heating/cooling effects reduces this difference only very marginally. We then conclude that the primary difference is driven by the switch to inhomogeneous \textit{X}-ray emission and energy deposition. The difference is comparable to, or below, the 20\% expected precision of semi-numerical simulations like \cmfast~\cite{Zahn:2010yw}.

\subsection{Tests of Energy Injection against \cmfast}

\begin{figure}[!t]
    \begin{center}
    \includegraphics[width=0.49\textwidth]{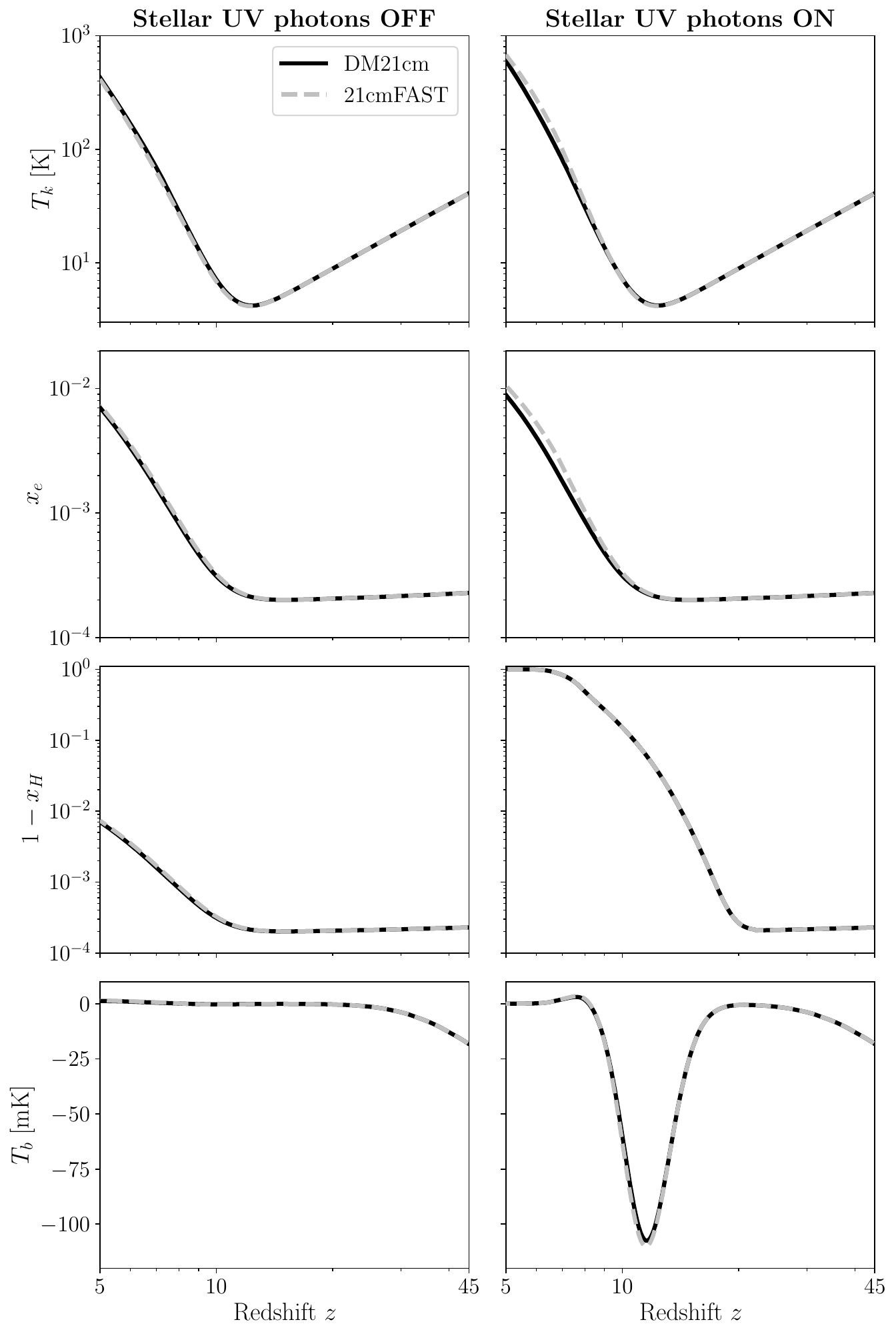}
    \caption{\textbf{Comparison of \dmcm and \cmfast global evolution under star formation \xray injection.} (\textit{Left column}) Global evolutions of $T_k$, $x_e$, $x_\mathrm{HI}$, and $T_{21}$ comparing \xray treatments of \dmcm and \cmfast, with locally ionizing UV radiation disabled.
    (\textit{Right column}) Global evolutions with ionizing UV radiation enabled. Generally good agreement is observed. See text for details.
    }
    \label{fig:xc_xrayV21_nopop2}
    \end{center}
\end{figure}

We now proceed to test our inhomogeneous \textit{X}-ray treatment against that of \cmfast. First, we first perform runs with \dmcm and \cmfast in their default configuration with the exception that we disable ionization from UV emission. Our \dmcm run uses our custom implementation of the \textit{X}-rays from the SFRD while the \cmfast run is unmodified. This is a particularly nontrivial test as our \dmcm prescription for attenuation differs from that of \cmfast.
The results of this first test are shown in the left column of Fig.~\ref{fig:xc_xrayV21_nopop2}, where we find particularly good agreement between the two codes. The maximum relative discrepancies in the global quantities are 9\%, 6\%, and 5\% in $T_k$, $x_e$, and $T_{21}$ respectively. Note that since stellar UV photons have been turned off, the evolution of $x_\text{HI}$ is identified with that of $x_e$.

We now proceed to restore the UV emission associated with stars, which drives ionization and magnifies the effect of \textit{X}-ray attenuation. Results for the evolution of the global quantities are shown in the right panels of Fig.~\ref{fig:xc_xrayV21_nopop2}. Here, the differences in our attenuation procedure become more readily apparent, with the maximum relative error in the global $T_k$, $x_e$, $T_{21}$ reaching 20\%, 25\%, and 15\% respectively. We are not surprised by this difference, as our total photoionization cross sections and branching fraction to each deposition channels are calculated using an improved treatment in \dhis, compared to those from Ref.~\cite{Furlanetto:2009uf}, used in \cmfast.

\section{Tests of Energy Injection from Dark Matter}
\label{app:DM}

In addition to cross-checking energy injection from star formation \textit{X}-rays, we compare the global evolution of \dmcm and \dhis under DM energy injection as it will be the type of energy injection handled by \dmcm in an actual simulation run. We exclude astrophysical \xray and UV photons in the evolution to showcase the effects of DM injection.

As previously discussed, the evolution of \dmcm and \dhis differs in terms of their treatment of helium ionization and the DM and baryon inhomogeneities. The two codes also differ through numerical integration accuracy (Euler's method for \dmcm vs. Runge-Kutta for \dhis) and the finite precision of parameters entering the time step $\Delta t$ calculation inherited from \cmfast. Here, we would like to focus on discrepancies due to the slightly different interpolation scheme of secondary particle emission and energy deposition between \dmcm and \dhis. As described in detail in Sec.~\ref{sec:DMInj}, we obtain the secondary photon and deposition transfer functions for \dmcm by injecting test photons and electrons on a grid of redshifts, ionization fraction, baryon density, and injection energy. The outputs particle number for each bin, and injected energy in each channel are tabulated and linearly interpolated in a \dmcm run. \dhis, while relying on interpolating precomputed data tables itself, computes a subset of transfer functions on-the-fly. Specifically, for photons the deposition from photoionization, and for electrons the inverse Compton scattering (ICS) transfer functions are computed in real time, since these transfer functions can depend quite sensitively on the present ionization fractions of hydrogen and helium (see Sec. III D and Sec. F of Ref.~\cite{Liu:2019bbm}). In \dmcm, it would not be feasible to perform such a real-time calculation, since it would require a different set of transfer functions for each cell with essentially unique combinations of baryon density and ionization fraction, both of which affect the total scattering cross section strongly. We thus choose a interpolation grid described in Sec.~\ref{sec:DMInj} and interpolate the secondary photon numbers and injected energies. The grid size is constrained by the finite GPU memory, with the GPU-accelerated treatment of the interpolation tables representing the limiting factor for the evaluation speed of \dmcm.

\begin{figure}[h!]
    \begin{center}
    \includegraphics[width=0.49\textwidth]{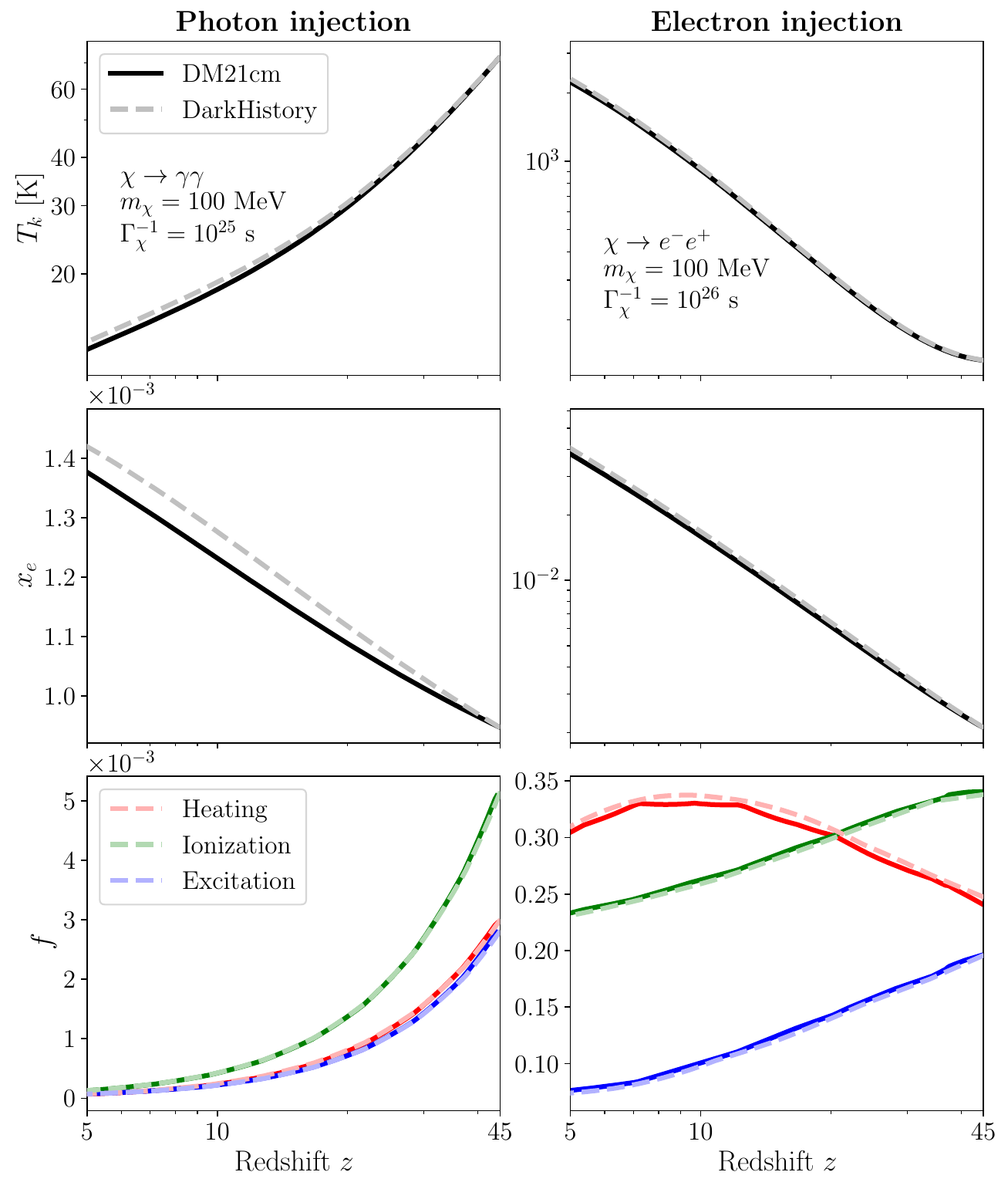}
    \caption{\textbf{Comparison of \dmcm and \dhis global evolution under DM energy injection.} For DM decaying into photon pairs (left column) and electron-positron pairs (right column), we compare $T_k$ (top row), $x_e$ (middle row) and energy deposition fraction (bottom row) calculated by \dmcm and \dhis in their default configuration, with star formation \xray and UV photons turned off in \dmcm. Notably, injection and deposition in \dmcm takes into account spatial inhomogeneity of DM. Differences in $T_k$ reach 5\% and 3.5\%; differences in $x_e$ reach 3\% and 6\% for photon and electron injection respectively, consistent with our understanding of their differences in helium treatment, numerical approximations, and spatial inhomogeneity. We also observe percent level discrepancy in the electron energy deposition ratio due to interpolation artifacts in the electron transfer function, a result of storage and memory constraint. See text for more details.}
    \label{fig:xc_inj}
    \end{center}
\end{figure}

Fig.~\ref{fig:xc_inj} compares the global evolution of \dmcm and \dhis under DM decaying to photons and electrons, for which the matter temperature $T_k$ and ionization fraction $x_e$ agrees generally at sub-10\% level. The figure additionally shows $f$, a dimensionless quantity defined as the energy deposited in a particular channel, normalized by the energy injected in the time step (same definition as in \dhis). While the photon deposition $f$'s are precise at the percent level, the electron deposition $f$ agrees with \dhis's up to 3\% due to the finite interpolation precision and rapid change of deposition behavior with the ionization fraction.

\section{Test of Field-level Information in \xray Energy Injection}
\label{app:PressSchech}
In this section, we examine the compatibility of our \xray deposition treatment for DM energy injection with the \xray deposition procedure in \cmfast. This is important as \cmfast models energy deposition following the formalism of Ref.~\cite{Furlanetto:2009uf}, while we make use of the more detailed \dhis approach~\cite{Liu:2019bbm}. By showing that \cmfast through its implementation of \cite{Furlanetto:2009uf} and our energy deposition reproduce similar spatial morphologies in the 21-cm signal when modeling the same energy injection, we confirm that our projected limits are robust with respect to systematic modeling differences between the two codes. 

We first expand on Sec.~\ref{sec:XrayComparison} on the \xray deposition treatment of \cmfast. As with \dmcm, \cmfast assumes \xray photons are emitted isotropically, and calculates the incident \xray at any given point at redshift and location $(z, \vx)$ by integrating shells at each past redshift $z'$ on the past lightcone. Instead of tracking an emission history, \cmfast re-calculates the past \xray luminosity at each step of the simulation using a conditional Press-Schechter treatment and emissivity field that is normalized an unconditional Sheth-Tormen calculation to obtain the differential flux received by the simulation cell at $(z, \vx)$ from $z'$, $dF_X/dz'(z,z',\vx,E)$. Using the flux, \cmfast integrates the total photoionization cross section using the global averaged nucleus number density (for each species being ionized), and then uses the branching fractions from Ref.~\cite{Furlanetto:2009uf} to calculate the deposition into each channel in heating, ionization, and Ly$\alpha$ excitation. In doing so, \cmfast makes the following assumptions: (1) the \xray spectrum is characterized by a power law, $dN_X/dz'(z',\vx,E)\propto E^{-1}$, which simplifies the frequency integral and redshifting; (2) the attenuation due to energy deposition is implemented in a discrete, on/off manner, i.e., when the optical depth of a particular mode $\nu$ from $z$ to $z'$, $\tau(z,z',E)$ is greater than 1, the mode is discarded; (3) the scattering rates, temperature change, and ionization fraction change are calculated assuming a global averaged nucleus number density, from cross sections, heating deposition, and ionization deposition respectively. This assumption is accurate only in the optically thin limit. For more details, we refer the readers to Sec. 3.1.2 of Ref.~\cite{2011MNRAS.411..955M}.

To make the comparison to \cmfast as close as possible, we have modified \dmcm to (1) inject the same power law spectrum calculated from the same hybrid Press-Schechter Sheth-Tormen procedure; (2) track the energy dependent attenuation factor separately from the spectrum associated with each shell, and discard any energy bins for which the attenuation exceeds a factor of $e$; (3) calculate the injection and energy deposition assuming the local nucleus number density to be the global average. Additionally, we simplify the conditional Press-Schechter calculation of the collapsed fraction to $f_\text{coll}\propto1+\bar\delta^R$ in \textit{both} codes, where $\bar\delta^R$ is the averaged density contrast within a sphere of radius $R$.

\begin{figure}[h!]
    \begin{center}
    \includegraphics[width=0.49\textwidth]{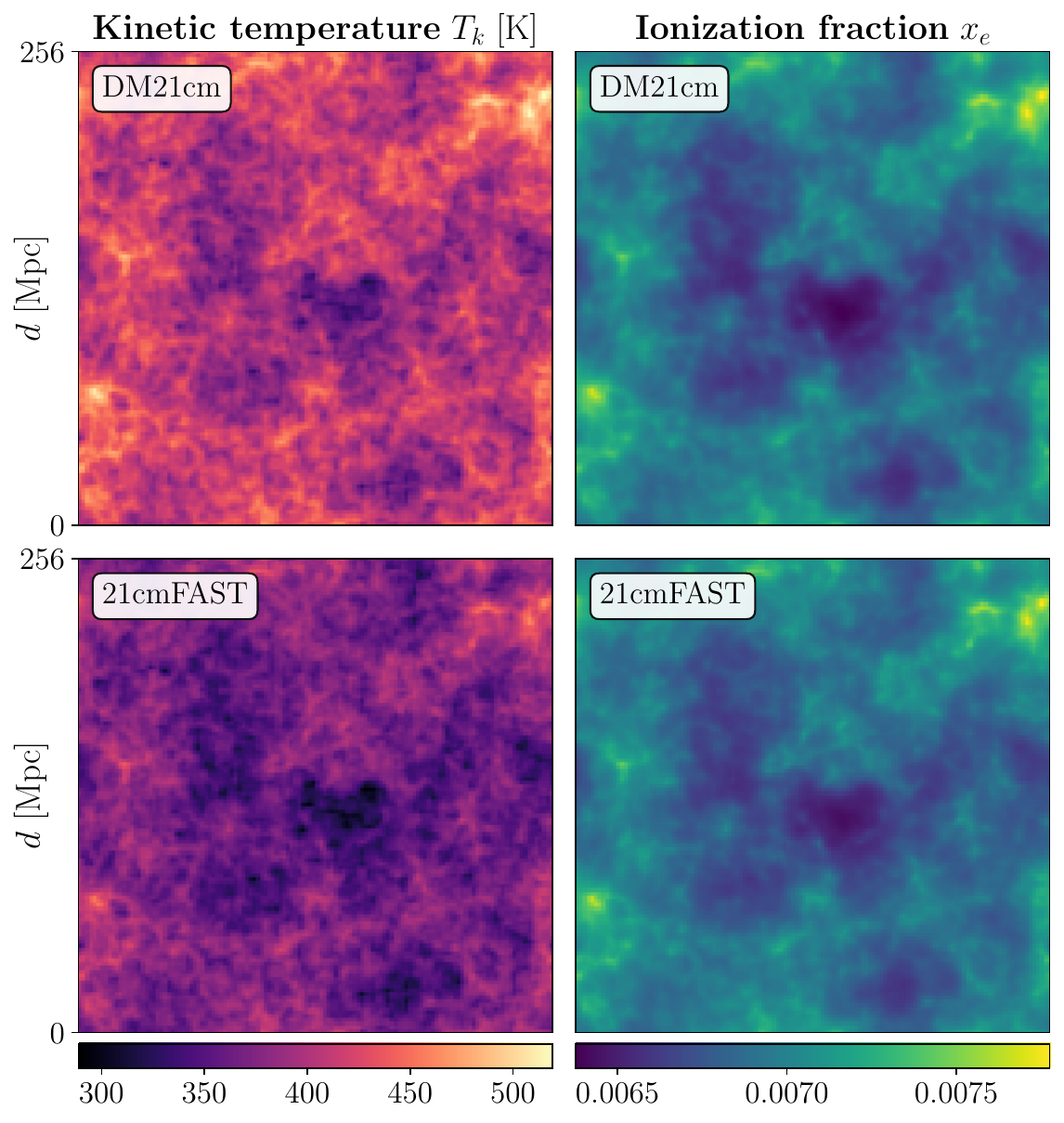}
    \caption{\textbf{Cross checking \dmcm against \cmfast's stellar \xray deposition.} The top row shows the $z=5$ $T_k$ (left column) and $x_e$ (right column) of the universe at the end of a \dmcm evolution, modified to match \cmfast's stellar \xray injection. The bottom row shows the default \cmfast result. Modifications were made to \dmcm to match \cmfast's assumptions for an instructive comparison. While displaying very similar spatial features, the \dmcm kinetic temperature has a slightly higher mean. This is due to a difference in the overall photoionization cross section and the fraction of energy deposited to heating, an effect also shown in Fig.~\ref{fig:xraycheck-ps}. See text for more details regarding the comparison.
    }
    \label{fig:xraycheck-field}
    \end{center}
\end{figure}

\begin{figure}[h!]
    \begin{center}
    \includegraphics[width=0.49\textwidth]{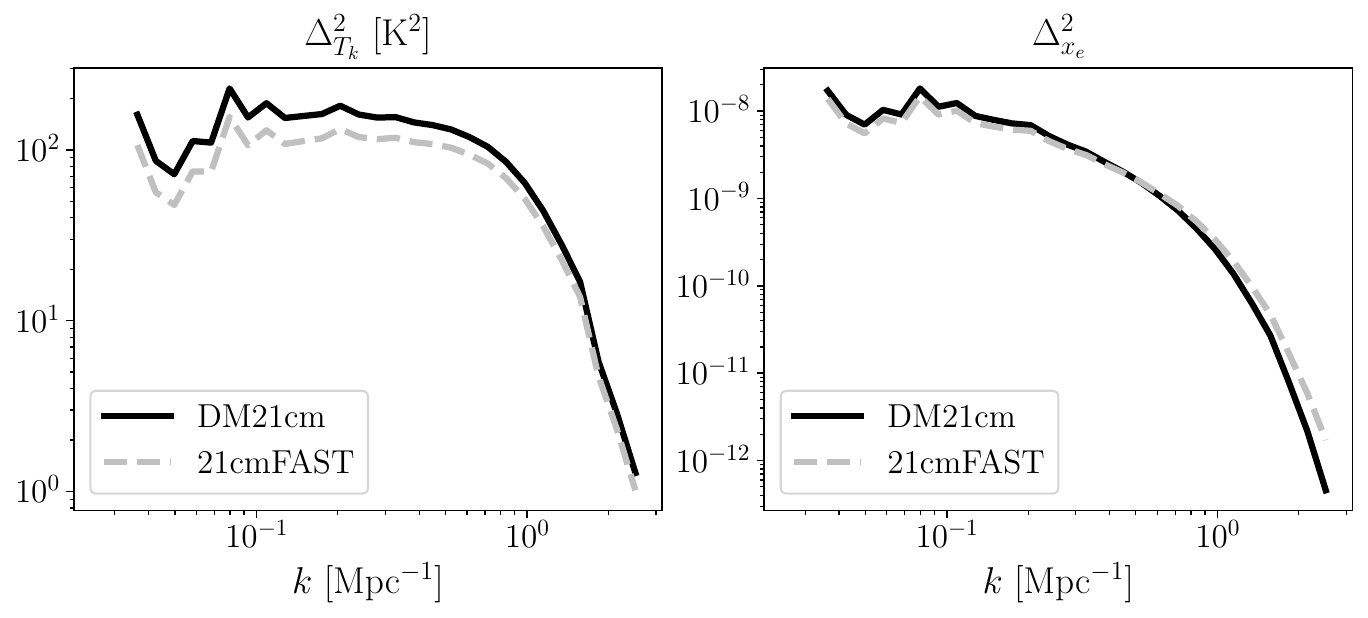}
    \caption{\textbf{Power spectra comparison between \dmcm and \cmfast's stellar \xray deposition.} We plot the power spectra of $T_k$ (left) and $x_e$ (right) for the $z=5$ final states shown in Fig.~\ref{fig:xraycheck-field}. The ionization fraction power spectrum $\Delta^2_{x_e}$ differs between the two implementation by a smooth factor of up to 12\% for $k<1$~Mpc$^{-1}$, while the kinetic temperature power spectrum $\Delta_{T_k}$ differs by a smooth factor of up to 24\%. This is consistent with the discrepancy observed in global signals originating from difference in scattering cross sections and branching fractions, with photons of different energy affecting their respective propagation length scale differently. See Fig.~\ref{fig:xraycheck-field} and text for more details.}
    \label{fig:xraycheck-ps}
    \end{center}
\end{figure}

Under these assumptions, we are able to obtain excellent agreement with \cmfast at field level for the evolved quantities kinetic temperature $T_k$ and ionization fraction $x_e$, as shown in Fig.~\ref{fig:xraycheck-field}. The power spectrum computed in these fields also agrees to a high precision, as shown in Fig.~\ref{fig:xraycheck-ps}. We note that the contribution from UV photons due to star formation have been turned off for this comparison, in order to showcase the relatively weaker ionizing effects of \xray photons. As noted earlier, \cmfast and \dmcm use slightly different photoionization cross sections and branching fractions to each channel, which is responsible for the overall shift in the kinetic temperature and its power spectrum.

\section{Spatiotemporal Resolution}
\label{app:SpatioTemporal}

\begin{figure*}[h!]
    \begin{center}
    \includegraphics[width=0.7\textwidth]{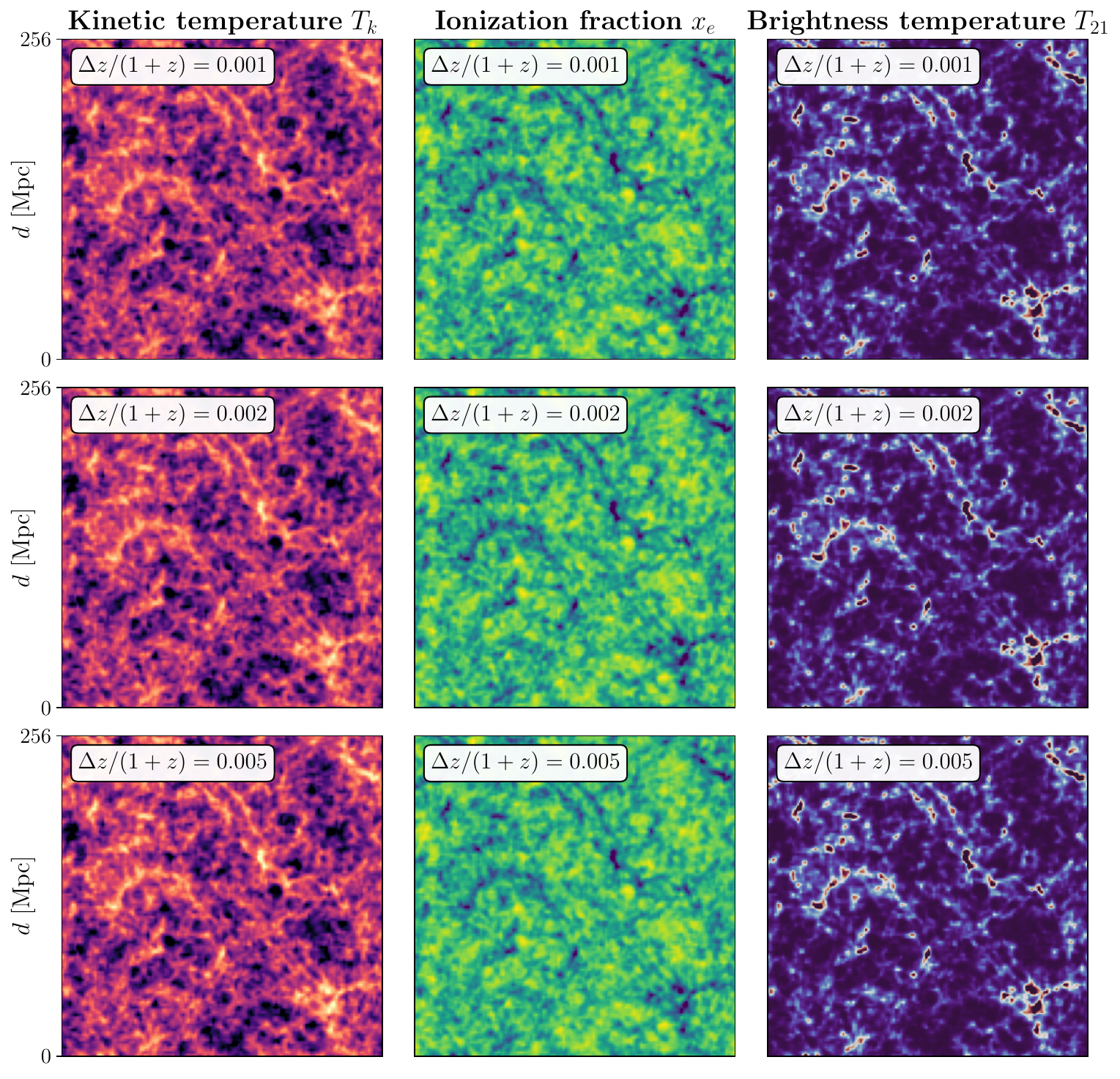}
    
    \caption{\textbf{Convergence of field features under DM energy injection.} We compare $T_k$, $x_e$, and $T_{21}$ field features evolved to $z=5$ under a strong DM injection scenario without astrophysical \xray and UV injections. The three rows are evolved under different \dmcm subcycle time steps $\Delta z/(1+z)=0.001,0.002,0.005$, with subcycling factors 20, 10, 4 respectively such that the \cmfast time step is $\Delta z/(1+z)\approx0.02$. By eye, the fields appear well converged. We quantify the $T_{21}$ convergence in Fig.~\ref{fig:convergence-ps}.}
    \label{fig:convergence-field}
    \end{center}
\end{figure*}

\begin{figure*}[h!]
    \begin{center}
    \includegraphics[width=\textwidth]{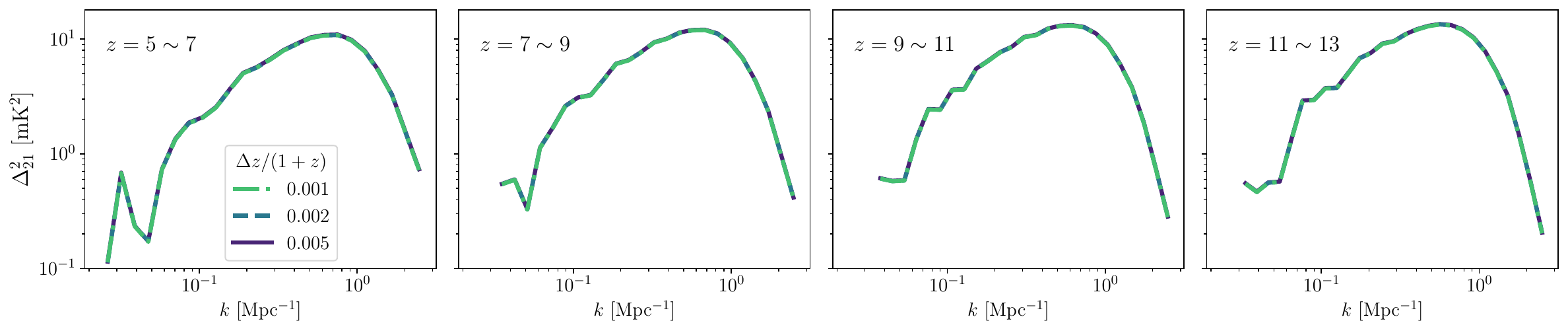}
    
    \caption{\textbf{Convergence of $T_{21}$ power spectra under DM energy injection.} Under the same setup as in Fig.~\ref{fig:convergence-field}, we show that the lightcone $T_{21}$ power spectrum is well converged as we decrease the \dmcm time step. We verified that the change of power spectrum is at 0.1\% level when decreasing the time step from $\Delta z/(1+z)=0.002$ to 0.001. We choose $\Delta z/(1+z)=0.002$ as our fiducial time step.}
    \label{fig:convergence-ps}
    \end{center}
\end{figure*}

In Fig.~\ref{fig:convergence-field} and Fig.~\ref{fig:convergence-ps}, we demonstrate the convergence of field level features in the kinetic temperature, ionization fraction, and 21-cm brightness temperature predicted by \dmcm. To showcase the effects of dark matter energy injection as opposed to standard astrophysics, we have turned off all star formation \xray and UV contributions. We chose a \xray injection scenario of a $m_\chi=3$~keV DM decaying to photons into the xray band with a large decay rate of $10^{-25}$~s$^{-1}$. The three evolutions shown are produced with $\Delta z/(1+z)=0.005,0.002,0.001$, with the subcycling factor being 4, 10, and 20 respectively, such that the \cmfast step size stays close to the fiducial $\Delta z/(1+z)=0.02$ value. (We have found \cmfast to be converged at the $\Delta z/(1+z)=0.02$ level, prior to \dmcm, due to the different \xray injection and deposition procedure. In particular, \cmfast can evaluate contributions of very recent past redshifts with large step sizes, which is important in producing the small scale features. \dmcm on the other hand relies on small step size in order to have access to information of the recent past.) We have found that the brightness temperature power spectra in all redshifts change by one part in $10^{3}$ when decreasing the time step from $\Delta z/(1+z)=0.002$ to $0.001$. For fast evaluation, we chose $\Delta z/(1+z)=0.002$ as our fiducial step size.

\section{Lightcones without Background Astrophysics}
\label{app:ExtendedLightCones}

\begin{figure*}[!t]  
    \hspace{0pt}
    \vspace{-0.2in}
    \begin{center}
    \includegraphics[width=0.99\textwidth]{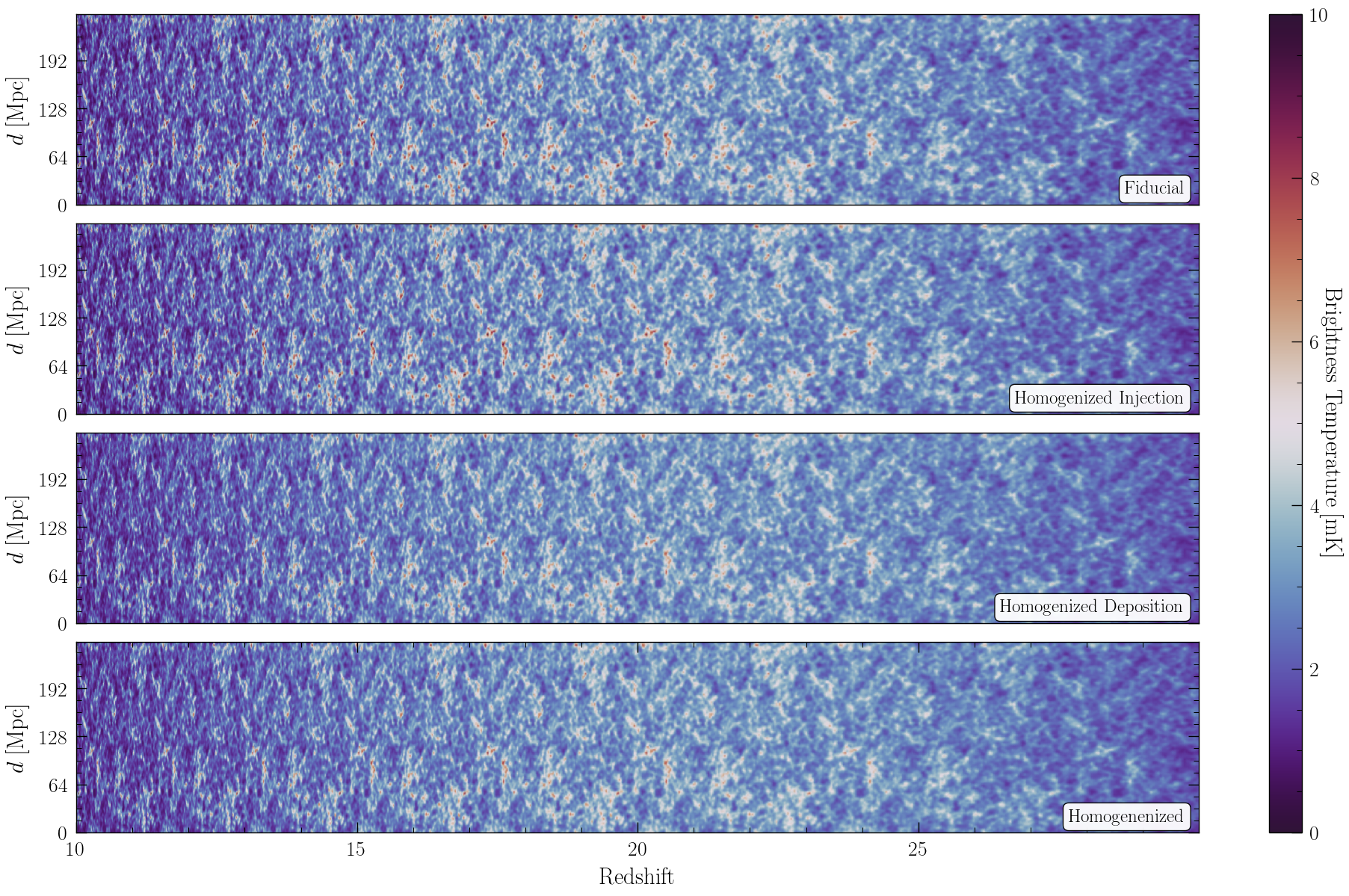}
    \caption{\textbf{Effects of homogenized injection/deposition on the $T_{21}$ lightcone under DM decaying to photons.} The brightness temperature lightcone evaluated using our fiducial simulation procedure and three systematic variations for DM with $m_\chi = 5\,\mathrm{keV}$ decaying monochromatically to two photons with a lifetime of $\tau = 10^{26}\,\mathrm{s}$, similar to Fig.~\ref{fig:Lightcones}. For illustrative purposes, the ionizing and heating effects of star formation have been excluded from this calculation. Comparing the results in which one or both of the energy emission and deposition are homogenized reveals that for photons, it is spatial inhomogeneity in the efficiency of energy deposition which most strongly determines the lightcone morphology as opposed to the inhomogeneity is injection, which is mitigated by the fact that 5~keV photon has a relatively long path length compared to electrons.
    }
    \label{fig:PhotonsNoAstro}
    \end{center}
\end{figure*}

\begin{figure*}  
    \hspace{0pt}
    \vspace{-0.2in}
    \begin{center}
    \includegraphics[width=0.99\textwidth]{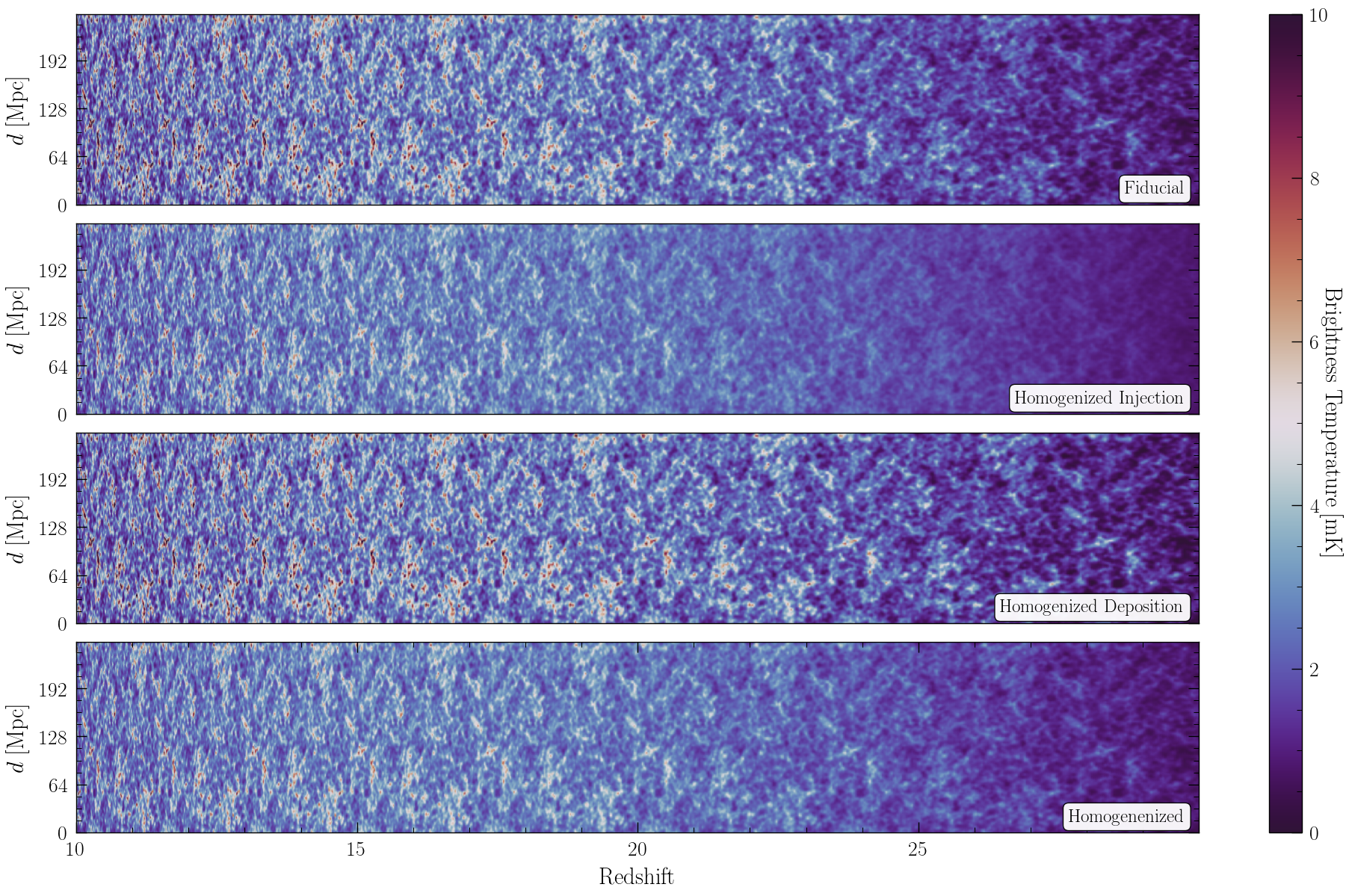}
    \caption{\textbf{Effects of homogenized injection/deposition on the $T_{21}$ lightcone under DM decaying to electrons.} As in Fig.~\ref{fig:PhotonsNoAstro}, but for DM decay to electrons for $m_\chi = 10\, \mathrm{MeV}$ and $\tau = 10^{25}\,\mathrm{s}$. Unlike in the scenario for decay to photons, we find that it is the spatial inhomogeneity in electron emission tracking the DM distribution which most strongly determines the lightcone morphology.}
    \label{fig:ElectronsNoAstro}
    \end{center}
\end{figure*}

In Fig.~\ref{fig:PhotonsNoAstro} and Fig.~\ref{fig:ElectronsNoAstro}, we provide example lightcones evaluated between $z=5$ and $z=30$ of the brightness temperature like in Fig.~\ref{fig:Lightcones}, but now with all background astrophysical processes typically modeled with \cmfast disabled. This allows for a cleaner identification of the effects in inhomogeneity in the photon and electron decay scenarios. We also provide intermediate results in which one but not both of the emission and injection processes are homogenized, allowing for inspection of the relevance of each aspect of inhomogeneity. In the case of decay to photons, it is the inhomogeneity in the energy deposition efficiency which primarily drives the lightcone morphology, while it is the opposite in the case of decay to electrons, where the inhomogeneity of the emission of decay products is most important.

\section{Expanded Triangle Plots}
\label{app:FullTriangle}

\begin{figure*}[!t]  
    \hspace{0pt}
    \vspace{-0.2in}
    \begin{center}
    \includegraphics[width=0.99\textwidth]{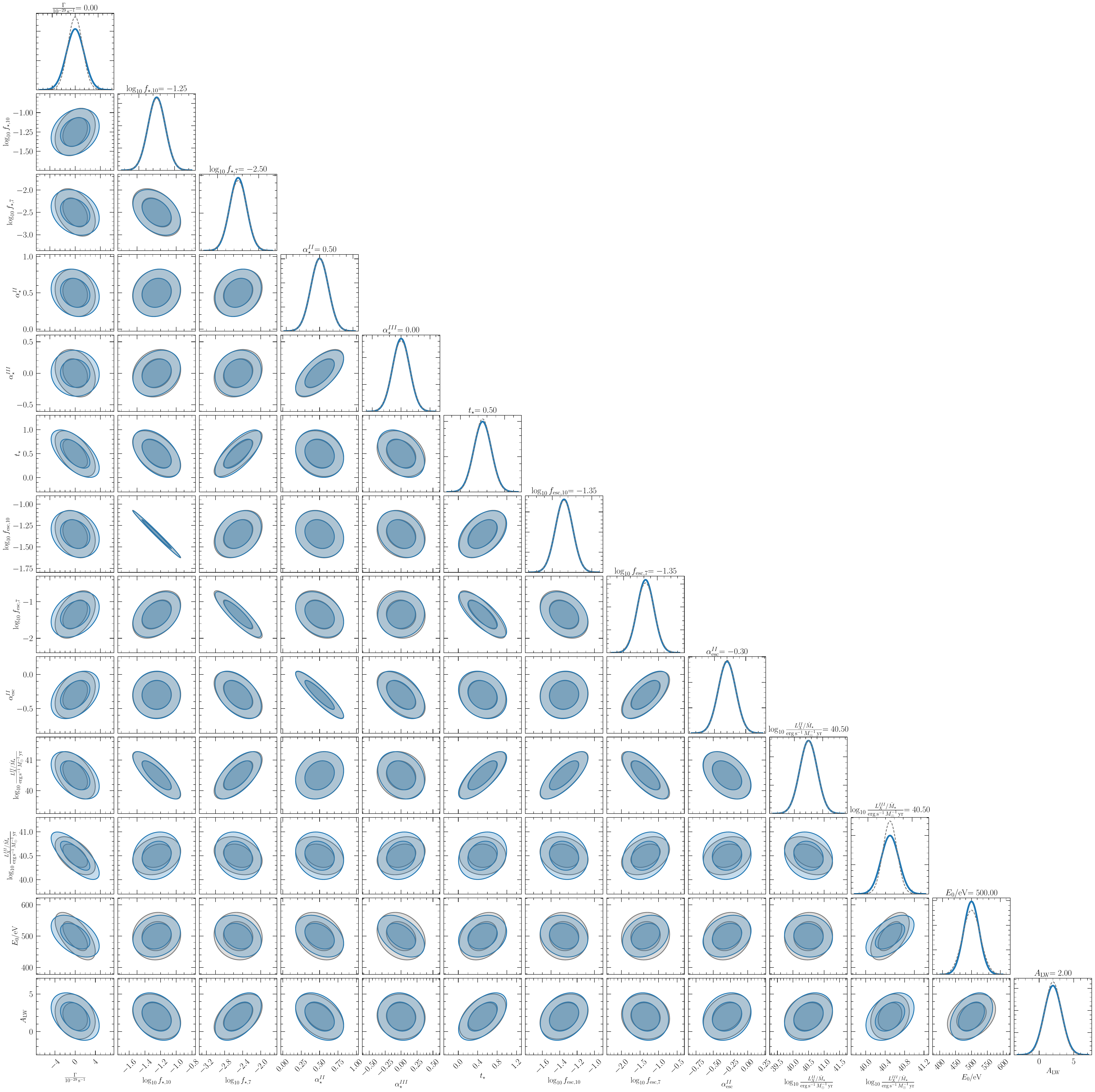}
    \caption{\textbf{Extended covariance between decay rate and astrophysical parameters in $\chi\rightarrow\gamma\gamma$.} As in Fig.~\ref{fig:PhotonTriangleSubset}, but now showing the full triangle plot with all parameter covariances depicted.}
    \label{fig:PhotonCornerFull}
    \end{center}
\end{figure*}

\begin{figure*}[!t]  
    \hspace{0pt}
    \vspace{-0.2in}
    \begin{center}
    \includegraphics[width=0.99\textwidth]{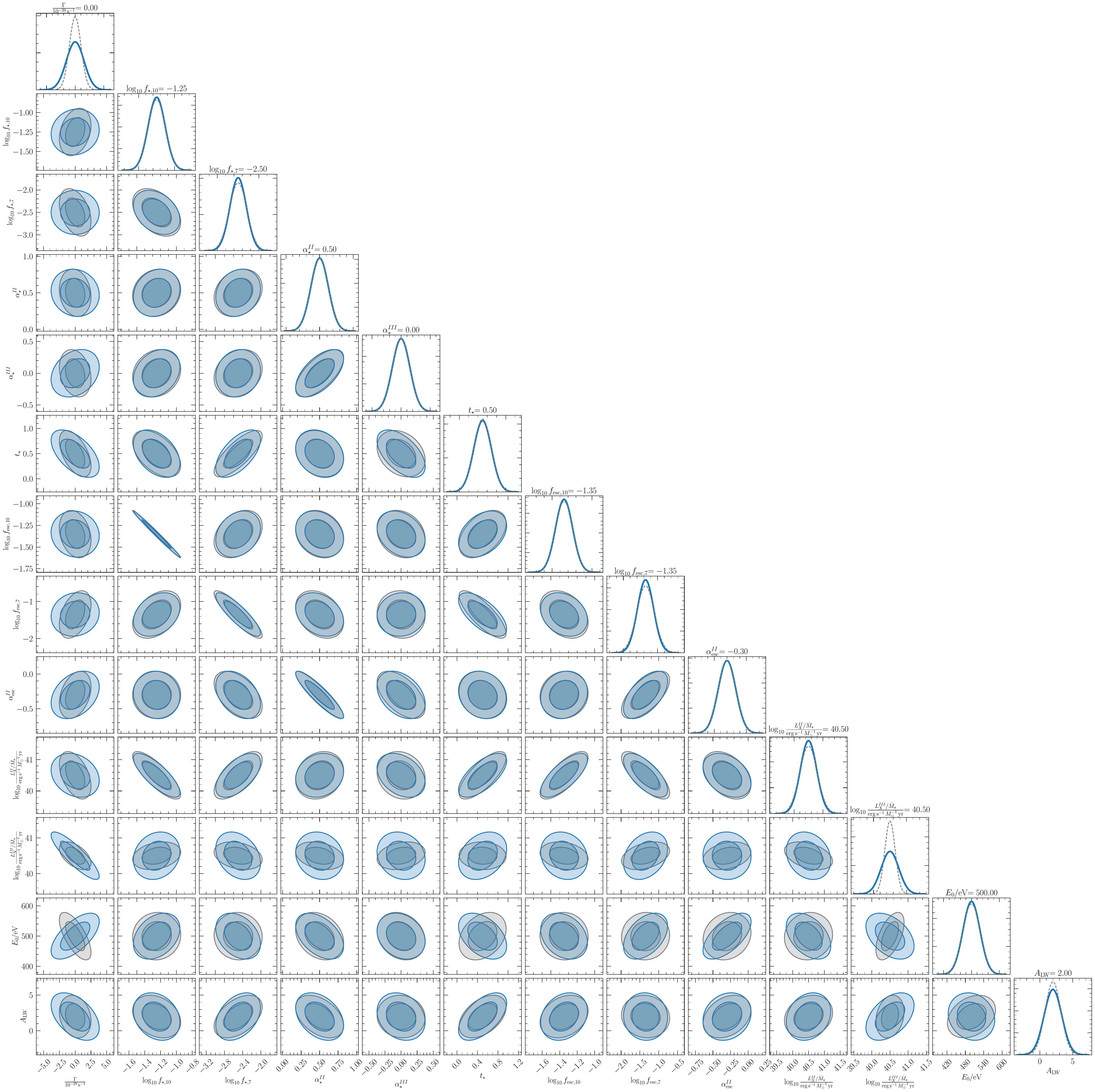}
     \caption{\textbf{Extended covariance between decay rate and astrophysical parameters in $\chi\rightarrow e^+e^-$.} As in Fig.~\ref{fig:ElectronTriangleSubset}, but now showing the full triangle plot with all parameter covariances depicted. }
    \label{fig:ElectronCornerFull}
    \end{center}
\end{figure*}

In this appendix, we present the full triangle plots for two representative scenarios of decay to photons for $m_\chi = 100\,\mathrm{eV}$ in Fig.~\ref{fig:PhotonCornerFull} and decay to electrons for $m_\chi = 100\,\mathrm{MeV}$ in Fig.~\ref{fig:ElectronCornerFull}. These results expand on those presented in the main text in Fig.~\ref{fig:PhotonTriangleSubset} and Fig.~\ref{fig:ElectronTriangleSubset}, respectively.
It is worth noting that the degeneracy direction between the \xray luminosity $L_X$ and the low-energy cutoff $E_0$ in Fig.~\ref{fig:PhotonCornerFull} switches when correctly modeling the DM energy deposition. This is to be expected, as a higher (lower) energy cutoff $E_0$ makes astrophysical deposition more (less) homogeneous.

\clearpage
\bibliography{main}

\end{document}